# A research roadmap for assessing the feasibility of warming Mars

E. S. Kite[1,2], A. Essunfeld[2,3], M. H. Hecht[4], M. A. Mischna[5], R. Wordsworth[6], H. Mohseni[7], A. Boies[8], N. Averesch[9], S. Ansari[7], M. I. Richardson[10], E. A. DeBenedictis[11], D. Stork[11], A. L. Bamba[7], C. J. Handmer[12], C. Jourdain[8], R. Ramirez[13], C. E. Mason[14], A. Kling[2], A. S. Braude[2], A. Dumitrescu[2], S. P. Worden[15], J. Cumbers[16], N. Lanza[3], R. Quayum[6], C. S. Cockell[17].

*1. University of Chicago (kite@uchicago.edu); 2. Astera Institute; 3. Los Alamos National Laboratory; 4. MIT Haystack Observatory; 5. JPL-Caltech; 6. Harvard University. 7. Northwestern University; 8. Stanford University; 9. University of Florida, Gainesville; 10. Aeolis Research; 11. Pioneer Labs; 12. Terraform Industries; 13. University of Central Florida; 14. Weill Cornell Medicine; 15. Breakthrough Initiatives; 16. SynBioBeta; 17. University of Edinburgh.*

## Summary

This roadmap outlines research pathways to determine whether Mars could be warmed with non-biological methods. It does not presuppose that warming Mars is desirable; its purpose is to identify what would need to be true for Mars to be warmed, what it would cost, and what could go wrong. Three complementary research tracks appear promising. Solid-state greenhouse membranes offer local warming, aiding water harvesting, food production, and oxygen supply near human bases. Orbiting reflectors can warm key sites such as bases and $CO_2$-ice reservoirs, although a large combined area would be required. Strengthening Mars' natural greenhouse effect might warm large regions or the globe, although many aspects remain to be worked out. Each approach carries scientific and technical risks that research must address. Near-term priorities are on-Earth testing of key parameters that will determine whether engineered aerosol warming is realistically possible, assessing whether exponential production of bioplastic habitats is possible, and designing at-Mars process experiments. In the near term, the research proposed here is closely aligned with and supports research needed to understand Mars' atmosphere and volatile evolution and hazards to human explorers. The main external uncertainty is whether or not launch costs continue to fall. This is early-stage research, and we discuss key near-term decision points, alternative pathways, and payoffs if research outcomes are negative. We also outline build-out pathways if research succeeds and demand exists. Relatively modest research investments would keep open the option of extending life beyond Earth as Mars' scientific exploration continues.

# Table of contents

# 1. Introduction

For Mars, possible human-involved futures include the Antarctica model (no long-term sustainable habitats), local terraforming (local habitats and ecosystems supporting permanent inhabitation), or regional-to-global environmental modification. Each has been explored [1-14], but each faces significant scientific open questions, and as of 2026 we do not know enough to make an informed choice. Choices are guided by existing plans and agreements. The Outer Space Treaty, ratified by all space-faring countries, states that "[t]he exploration and use of outer space, including [Mars], shall be carried out for the benefit and in the interests of all countries, irrespective of their degree of economic or scientific development, and shall be the province of all humankind" [4]. NASA's Mars Future Plan [5] seeks to serve and preserve a wide range of "intergenerational interests, values, and opportunities in interplanetary peace and prosperity […] A stewardship role centers on a spirit of cooperation." Previous US Federally-charged commissions include the 1986 National Commission on Space, which says (of Mars) "Long-term exponential growth into eventual permanent settlements should be the overarching goal" [6], and the 2009 Augustine Commission [7], which concluded "the ultimate goal of human exploration is to chart a path for human expansion into the solar system," and that a "move toward[s] permanent expansion of human civilization beyond the Earth" was most likely on Mars. A 2025 National Academies consensus report on the science strategy for the human exploration of Mars prioritizes on-Mars research needed "as part of an integrated ecosystem of plants, microbes, and animals" during initial human missions to inform future choices about using Mars' resources to sustain "permanent inhabitation" [8]. The report states that in situ resource utilization "will extend beyond basic survival needs to support fuel production and habitat expansion."

These reports and associated research highlight that humans on Mars will initially follow the Antarctica model, but also highlight the limitations of that model. Logistics is much more challenging on Mars than in Antarctica, and the need for in situ resource utilization is correspondingly greater [9]. Closing the gap between the aspiration of permanent human presence and the scientific question of whether the environment can support it will require research.

Recently, applied astrobiology—the study of what would be needed to create sustainable habitats and biospheres beyond Earth [10]—has emerged as a new field of study. However, there is no recent systematic roadmap for the research needed to create the physical boundary conditions to support large numbers of people living on Mars.

Any deployment to extend life beyond Earth would involve many questions about the long term goals of space exploration (§1.1) [6-7,9,11]. The necessary (but insufficient) contribution of the research outlined in this roadmap is to better quantify how proposed actions map to costs, benefits, risks, and uncertainties. A finding that warming is not feasible would itself be valuable, as it would constrain planning for long-term human presence beyond Earth.

NASA published the first roadmap for Mars-warming research 50 years ago [11-13]. The European Space Agency carried out the first 3D modeling of an engineered Mars climate nearly 15 years ago [14]. However, the high cost of access to space and limited knowledge about Mars made further detailed study appear unpromising until recently. Several developments motivate a fresh look: falling launch costs, new evidence for the robustness of simple life [15], better spacecraft constraints on Mars' resources for biology and challenges to biology [16-19], and new warming ideas and climate models (e.g., [20]). It is necessary to think about what warming endpoints are achievable, whether they are desirable, and their associated costs, risks and benefits.

Whether Mars can support a biosphere is unknown. A biosphere on Mars would help sustain large numbers of people in bases beyond Earth [11-13,21-23], and set the initial conditions for a centuries-long process of atmospheric oxygen build-up. Earth's ecosystems are the foundation for a self-sustaining civilization. Biology, unlike machines, can complete assembly from base elements and grow exponentially without intervention—but only under favorable conditions, which Mars' surface currently lacks. We do not yet know enough to create a biosphere from scratch. Applied astrobiology, like planetary science, requires contributions from biology, chemistry, materials science, and climate modeling, among other disciplines. Life requires liquid water, and Mars is currently too cold for sustained near-surface liquid water. Warming Mars enough to melt ice is necessary but insufficient for a biosphere [20,24]. For example, Mars' soil contains perchlorates [18] and is rich in salts. Moreover, no known technology can quickly make Mars' atmosphere breathable.

This roadmap covers non-biological research. A parallel roadmap for the biological research needed to establish a biosphere is not yet available but is equally critical [22,25-27]. Relevant topics would need to include pioneer organism selection and engineering [25], biomaterials production, design of habitats capable of exponential production [28,29], pedogenesis [30,31], oxygen productivity across ecosystem types, biogeochemical feedbacks [32], and ecology [22,32-34]. Warming research does not invoke planetary protection concerns beyond those for any other uncrewed payload to Mars; biology experiments (such as the Italian Space Agency's plant growth experiment selected[1] to fly on a first SpaceX Mars landing) would be a separate decision. Key biological decision points (choice of organisms, inoculation go/no-go, target ecosystem choice) would need to be coupled to warming milestones and would likely best be developed in parallel. Introducing biology to bare soil outside protective membranes would likely be impractical until Mars were globally warmed, which is at least decades in the future. Early ecosystems would require monitoring of carbon, nitrogen and water cycling, with goals of retaining nitrogen and water in the biosphere, and burying carbon in the subsurface. These long lead times would provide an opportunity for supporting research and policy deliberation on what would be an important ethical as well as technical step (§1.1).

---

[1] Italian Space Agency, "Italy Goes to Mars," https://www.asi.it/2025/08/litalia-va-su-marte/

Warming methods can be evaluated against criteria including: compatibility with planned in situ resource utilization and biosphere tests [7], benefit to early bases, cost scaling, interaction with natural climate feedbacks, human health and environmental safety, and sustainable resource use. No method has been shown to clear this high bar; the research outlined here would determine which methods, if any, can do so [3,10-11,21].

Recent research includes demonstrations of photosynthesizer growth inside bioplastic habitats in Mars chambers [29], fabrication of candidate nanoparticles with infrared absorption matching predictions (H. Mohseni group at Northwestern University; Fig. 12), climate modeling of a warmed Mars [20,24,36], new warming-mechanism ideas [24,28,35], preliminary technoeconomic analysis (e.g., [37]), solar-sail orbital calculations, workshops (e.g., [10,38]), and addition of suitable-for-Mars traits into microorganisms.[2] Nevertheless, Mars-warming research currently occupies a tiny fraction of the total space research effort. The questions raised by the possibility of warming Mars are large, but the immediate unanswered questions are identifiable and can be addressed with a focused research campaign—which is what this roadmap attempts to lay out.

## 1.1 Ethics and planetary protection

Throughout the Space Age, the international Committee on Space Research (COSPAR) has made recommendations to reflect evolving scientific understanding of planetary protection, including planet categorization, organics and spore upper limits, and reporting of mission activities. These recommendations are, in practice, followed by all major space organizations.

There is no consensus about whether humans should extend life beyond Earth, and there is general agreement that the scientific search for existing life should occur first. Ethical objections include the possibility that life has already established itself on Mars but in places or in quantities that will be hard to recognize using scientific instruments [39], that Mars as it is—a desert world shaped by four billion years of geological processes—has intrinsic value that would be diminished or lost through human intervention [40], and that off-world work would be a misallocation of attention or resources [41]. Ethical arguments in favor of extending life beyond Earth include utilitarian and intrinsic worth considerations [42,43], creative-ecological arguments [44], the shift in perspective on life in the Universe that comes with expanding beyond Earth [10], and the idea that living worlds are intrinsically more valuable than dead ones [45,46].

For greening Mars, ref. [47] suggests three principles to guide research: 1. "[D]o more to show that life is absent on other worlds before we consider making them more habitable to life from Earth. [...] Most of the remaining possible locations for life on Mars could be searched via robotic missions in the price range of a few billion dollars." 2. "[K]eep learning about the solar system. Mars [still has] much to teach us about habitability and the origin of life on Earth, and we must be careful not to erase this archive until we have deciphered it." 3. "[S]ustainability: Greening efforts must not irreversibly consume any resources that might be vital to future generations. [...] Humanity's track record of environmental modification on Earth is all the more reason to proceed thoughtfully."

---

[2] Specifically, salt tolerance: https://pioneerlabs.substack.com/p/the-first-turn-of-our-engineering

Consensus requires more data on two fronts: whether Mars could support life in the future, and whether life endures on Mars today. In this context, specific existing priorities for planetary science, such as sample return and volatile mapping, are especially important (§4.1). Mars shows how a once-habitable planet can change climate over geologic time and lose its habitability: advocates of renewed habitability can learn much from these records.

All countries are committed to avoid harmful contamination of space [4,48]. According to the NASA Planetary Protection handbook [49], "The interpretation of harmful contamination changes as exploration technology evolves and scientific studies provide more knowledge. This flexibility is one reason why the Outer Space Treaty has stood the test of time [...]" Mars appears dead but was once habitable. If life established itself, adapted to worsening conditions, and was able to migrate to refugia, and refugia still exist on Mars, then life might persist in refugia. If scientific studies show that there is life on Mars, people may be much less likely to want to add Earth-derived life to the planet. Conversely, consensus that there is no life on Mars would make some people more likely to opt for making Mars more suitable for Earth-derived life. A thorough life search requires looking in more places, drilling deeper, and returning samples, and will be accelerated by human exploration [8]. Short-term, microbes from future human missions will stay within habitats, and although some will escape (as we have learned from the International Space Station) they are not expected to grow on Mars unshielded [48,50]. Long-term, planetary protection's key role will be determining whether and when the surface has been sufficiently searched that releasing terrestrial microbes into large warmed environments would not be harmful.

Governance questions are important as human numbers on Mars increase, but are not the focus of this roadmap. There are a wide range of perspectives on how human activity on Mars should be governed (or self-governed, e.g., [51]). Under current law, human activity in space would fall under the jurisdiction of the launching nations [4]. Local warming and small-scale tests could proceed under existing space law, though that is not the only basis which would need to be considered. Regional-or-larger climate changes would almost certainly invoke international coordination including agreement among spacefaring nations on the best scientific path forward. Relevant precedents include the International Seabed Authority, the Antarctic Treaty (Convention on the Regulation of Antarctic Mineral Resource Activities / Madrid Protocol), the Artemis Accords, COSPAR, the International Mars Exploration Working Group, and the Outer Space Treaty itself.

An approach to warming Mars would be inherently modular in that it could be done by many sites in parallel. As for many large completed scientific and technological projects, it invites international cooperation.

## 1.2. Physical constraints on Mars warming

Mars is dry because its water ice is too cold to melt. Warming it requires either increasing absorbed sunlight or strengthening the greenhouse effect. Increasing $CO_2$ atmospheric pressure provides modest additional warming, and widens the temperature range for which liquid water can occur.

Mars is warmed by absorbed sunlight and cooled by infrared radiation to space [52] (Fig. 1). These fluxes balance at ~$1.6 \times 10^{16}$ W, so Mars' climate (annual average 210 K) is stable on human timescales. Without an ocean or thick atmosphere, Mars responds in <1 year to changes in the balance of energy fluxes (Appendix A).

In 1973, Sagan proposed darkening Mars' surface [23], which can warm the surface by about 10 K (210 K→220 K). Sagan thought additional warming from climate feedbacks would be needed [23,53]. Models predict positive climate feedback could have boosted temperatures by up to 50 K regionally in Mars' past [54-57]. However, this invokes a mechanism—water-ice cloud greenhouse warming—that is not proven to have provided such strong warming in Mars' past [58], and a previously suggested $CO_2$-outgassing tipping point by itself does not sustain warmth [59,60]. Additional warming agents are needed to supplement Martian $CO_2$, as despite recent discoveries of carbonates [61,101], readily available $CO_2$ reserves (ice and adsorbed $CO_2$) are insufficient [59]. Taking account of this, none of the warming methods discussed here rely on the greenhouse effect of mobilized $CO_2$. We assume ongoing warming inputs would be needed (no tipping points), revisiting this in §6.

To melt ice requires either reflecting more sunlight onto the surface [13] or augmenting the greenhouse effect. Specifically, to create a warm season over large areas of Mars (surface temperatures above the freezing point averaged over a 70-sol period) we would need to supply at least $\sigma( (210 + 35)^4 - 210^4)A_{Mars} = 1.3 \times 10^{16}$ W, enough to raise Mars thermal emission by the blackbody equivalent of 35 K. This is not a strict requirement for biology (life can grow in brine below 273 K), but a rough proxy for the warming required to yield a warm season.

Gases (e.g., $CO_2$) and aerosols (e.g., clouds) can both warm surfaces by trapping or scattering thermal infrared radiation (Fig. 2) [56,62]. However, most natural aerosols cool Mars' dayside, reflecting more energy than they trap. Climate models are needed to assess net warming [52,63]. For engineered aerosols, the minimum atmospheric mass needed to change climate would be approximately 3 million tonnes (Appendix A) [64]. The actual value depends on the spectrum of the specific particle considered. On-Mars production of warming agents would therefore likely be the preferred approach. However, on-Mars manufacturing at scale is unproven and would require extensive de-risking. Assessing whether making engineered aerosol on Mars might be practical and feasible is a focus of the engineered aerosol research effort.

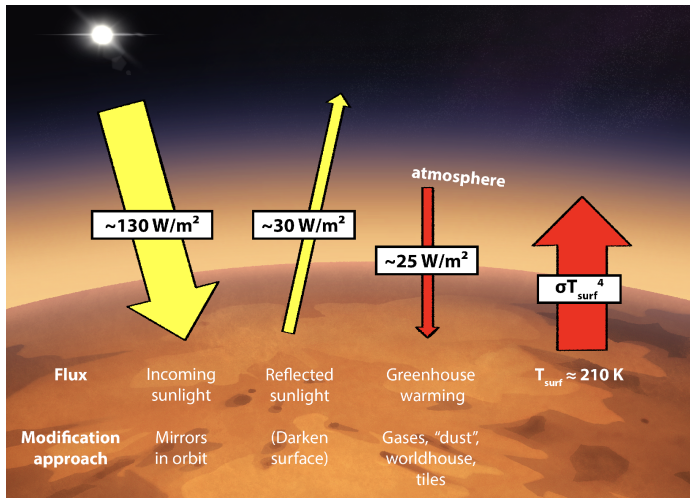

Fig. 1. (a) Energy fluxes that set Mars' temperature today, and potential Mars-modifying approaches that could increase sunlight or strengthen the natural greenhouse effect. At present, the net absorbed energy is 125 W/m$^2$, resulting in a surface temperature of $T_{surf} \approx$ 210 K. From [21]. Figure by D. Zhou. (b) Simplified schematic of three Mars-warming methods. Mars can be warmed using a space heater (orbiting reflectors), or an insulating blanket made either of solids (membrane warming approach) or aerosol in the atmosphere.

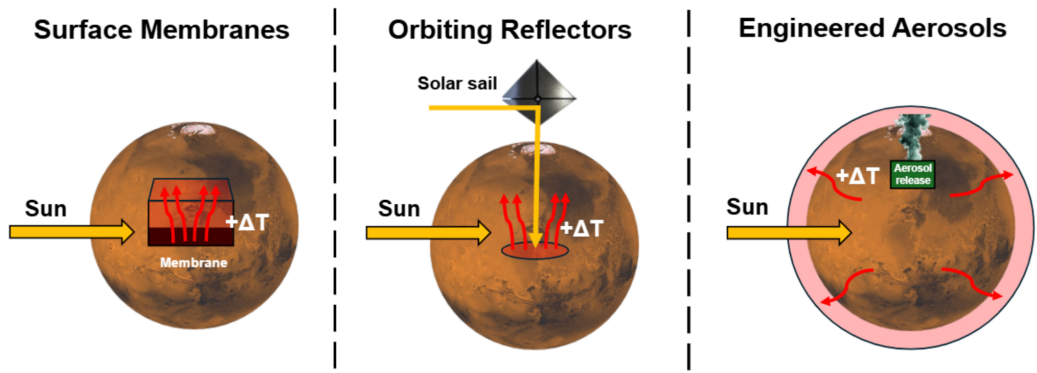

Local warming using solid-state greenhouse materials can provide benefits without global climate change [28]. Local methods would need to counter net loss of $H_2O$ ice to sublimation when warmed [71]. Local mitigation options include gas-tight membranes, condensation traps, or selectively permeable membranes. For practical purposes evaporative cooling prevents surface-exposed liquid water [72].

This document focuses on non-gas warming approaches, for the following reasons. Greenhouse gases warm Venus (+550 K), Earth (+35 K), and Mars (+5 K). However, warming Mars with artificial greenhouse gases would require >10$^{14}$ kg of fluorocarbons [65], and no adequate fluorine source is known on Mars. This is 10$^4$× more mass in the atmosphere than for aerosols [20]. Chlorine-based alternative warming gases are toxic to humans at small concentrations. Though finding alternative artificial greenhouse gases could be a topic for future research, and mass-in-the-atmosphere is only one of many criteria for comparing warming approaches, this is unpromising. It is easier to work with the materials that are already on Mars rather than import raw materials from beyond Mars. Moreover, the power demand of bringing volatiles such as $CO_2$ or $NH_3$ to Mars by redirecting asteroids is about half that of Earth's current total global power consumption (Appendix A); this does not appear to be a promising near-term direction [37].

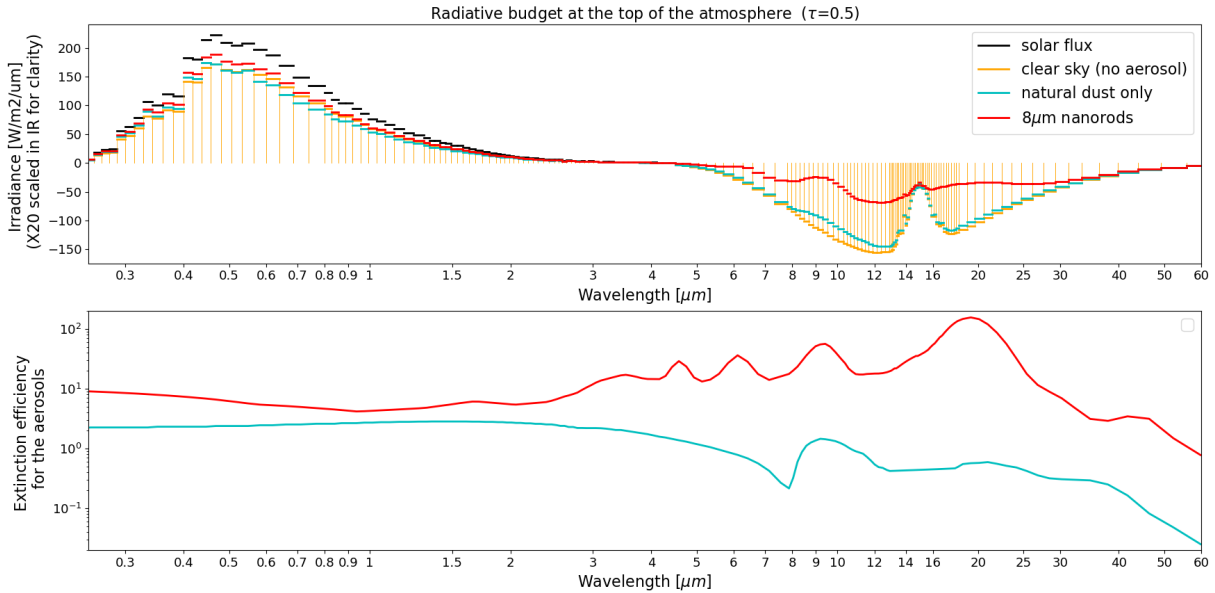

*Fig. 2. Top panel shows the solar energy input (short wavelengths) and thermal infrared energy losses (long wavelengths) at the top of the atmosphere, in equilibrium under standard dust conditions (cyan curve). Infrared absorption is weak everywhere except near 15 μm ($CO_2$ absorption). Energy balance shown for modern Mars' surface temperature. The instantaneous introduction of Mars-warming particles—here, engineered aerosol modeled in [20]—suppresses infrared cooling (red curve), which would result in a new equilibrium state at a warmer surface temperature (not shown). The bottom panel shows extinction efficiency for the natural (cyan curve) and engineered (red curve) aerosols. Natural dust has low extinction at long wavelengths, whereas the engineered particles have a 10× ratio of peak infrared to visible extinction. τ is optical depth (extinction) at 0.67 μm. Figure from TerraScreen[3].*

In the longer term, gases could return to the fore. Earth's habitability over geologic time is thought to have been maintained by negative feedbacks involving $CO_2$ and/or $CH_4$, in which biology plays a central role [73]. If a biosphere is established on Mars, then there are several possible biogenic gases that might help to maintain warm conditions.

---

[3] https://github.com/mars-terraforming-research/TerraScreen

**Box 1. Climate model improvements needed.** Mars terraforming climate models have been run by NASA, the European Space Agency, nonprofits, and academic scientists, with models including MarsWRF, the Kasting/Ramirez code, the Laboratoire de Météorologie Dynamique Global Climate Model (GCM), and the NASA Ames Mars GCM in 1D mode[3] [14,20,24,36,65-66]. Recent results suggest radiative-dynamical feedbacks are manageable [20] and suggest that warming will not destabilize ground ice [36], while highlighting that global water-cycle changes (Fig. 3) could take decades or more to reverse.

Needed improvements include better models of particle charging, dust cycle feedbacks, surface deposition and re-lofting, and microphysical interactions between natural and engineered aerosols (for instance, the ability for engineered aerosols to nucleate cloud particles) [67]. Key parameters (e.g., the contact angle between engineered aerosol and $H_2O$) require laboratory data. Model intercomparison studies (e.g., CUISINES [68]) are needed.

Long-term endpoint analysis would need to include $O_2$ as a major atmospheric component, allow estimates of biological productivity (food and oxygen) and the locations of lakes, and determine $CO_2$ needs for a stable warm climate.

Even sophisticated climate models can be wrong [69], and doubling energy input to a planet produces a wide range of outcomes. Small-scale, reversible experiments would need to precede any full-scale warming. Ongoing monitoring of Mars' natural climate variations [70] is needed.

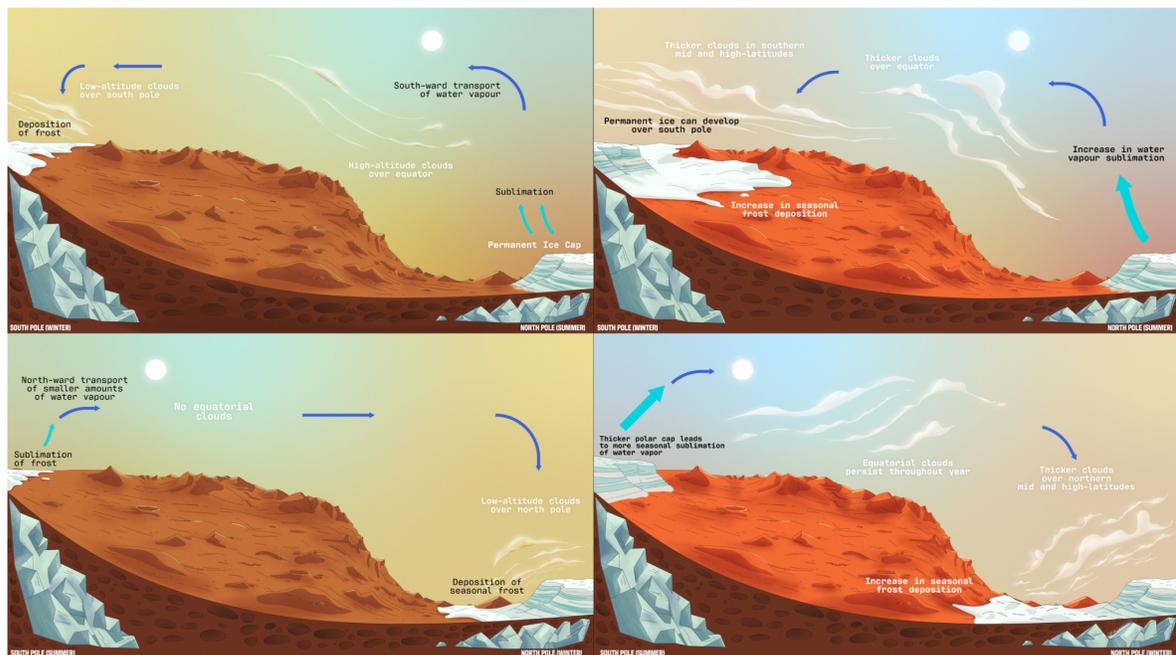

*Fig. 3. Differences in Mars' water cycle between the present-day Mars case (left) and the artificially warmed case (right) [36]. Cloud cover increases, and the South Polar water ice cap's role in the water cycle strengthens. Figure by V. Socianu.*

## 2. Overview of candidate Mars-warming approaches

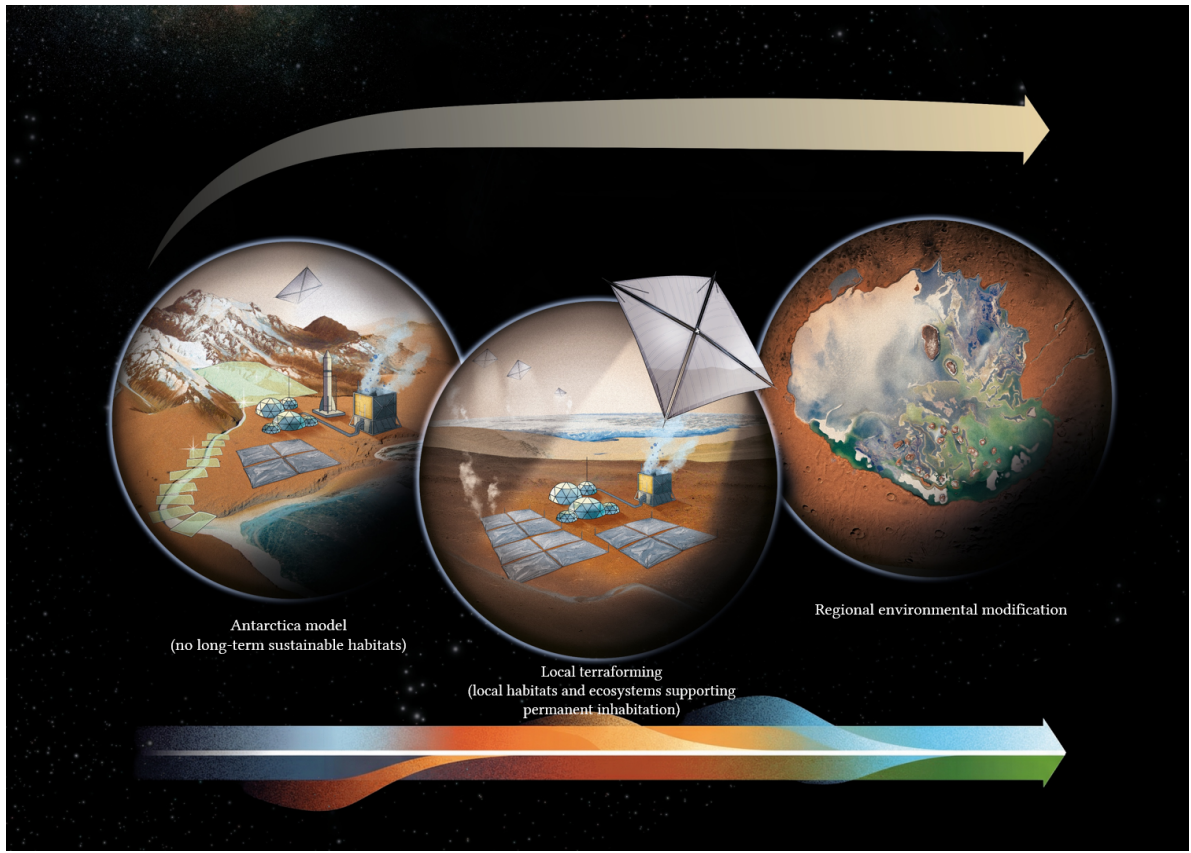

*Fig. 4. Artist's impression showing possible choices for the scale of human involvement in the future of Mars. Early on, warming occurs only close to human bases. Solid-state greenhouse membranes harvest liquid water from subsurface ice. Orbiting reflectors augment sunlight, and a pilot factory makes engineered aerosols. It is possible that people will decide to stop at this point. In the case where a decision is made to go for local terraforming, the middle figure shows atmospheric pressure rising from orbiting reflectors, larger areas with solid-state greenhouse membranes, and climate experiments with engineered aerosols. The right panel shows a large region (Hellas basin) suitable for photosynthetic life, in the case where a decision is made to proceed to regional warming. (V. Socianu). Research tasks to assess the feasibility of these options are summarized in Table 1.*

First steps towards assessing the technical viability of warming Mars include tests of local warming, including search-for-life experiments [7]. Any warming test program must be reversible at each stage, so that further warming can be halted if results are negative or costs are prohibitive (Fig. 4). Three ideas for non-biological warming of Mars have complementary strengths:

- Greenhouse membranes enable relatively inexpensive local habitats by decoupling the habitats for hardy simple organisms from the habitats for humans. However, they do not scale to large fractions of Mars' surface area unless they could be made from Mars materials.

- Orbiting reflectors work best for warming intermediate-sized areas ($10^2$-$10^5$ km$^2$), with a possible first application being illuminating human bases.
- Greenhouse agents (such as engineered aerosols) spread globally unless made deliberately short-lived, enabling process experiments to test models of climate feedbacks [14,20,36], but there are significant open questions about particle lifetime.

Which warming method is best depends on the desired human-involved future for Mars (Fig. 5). For a small population (~5,000, like Antarctica), membranes and possibly reflectors would likely suffice. For global warming, greenhouse agents appear most cost-effective.

To evaluate approaches we assume average transport costs to Earth orbit and to Mars' surface of \$100/kg and \$2,000/kg, respectively [74]. These costs are well below current prices: advertised prices are ~\$1,000/kg to Earth orbit and \$100,000/kg to Mars' surface. However, much lower costs are achievable with in-space propellant transfer and reusable vehicles currently undergoing test[4]. We assume that spacecraft going to Mars never return, so that every shipment to Mars must pay for the single-use Mars transport spacecraft. However, propellant for the voyage is delivered to Earth orbit by reusable spacecraft. Sensitivity to these assumptions is shown in Table 7. If launch costs do not fall below current values, that would favor strategic investment in on-Mars production of factory components in order to minimize the mass that would have to be transported from Earth (§7.2). Taking as an example a cost cap of \$1 bn/K/yr for +35 K warming (Appendix C), transport costs imply <18,000 tonnes to the Mars surface per year. Shipping dominates total cost because manufacturing costs are <\$2,000/kg. Shipping costs might fall below our assumed value if there is a propulsion breakthrough (such as nuclear-electric propulsion), or if Mars spacecraft are used more than once. Nevertheless, shipping costs are high and minimizing transported mass is therefore a focus of this roadmap.

$$\text{Cost} = \underbrace{\text{shipping}}_{\text{power, factory, consumables}} + \underbrace{\text{manufacturing}}_{\text{power, factory, consumables}} + \text{R\&D}$$

The cost of shipping power infrastructure to operate on-Mars factories can dominate overall cost. We bracket this uncertainty by considering an already-implemented, expensive option

---

[4] Companies with reusable launch vehicles currently undergoing test include Blue Origin, SpaceX, Stoke Space, LandSpace (蓝箭航天), Relativity Space, CAS Space (中科宇航), Rocket Lab, Galactic Energy (星河动力), Space Pioneer (天兵科技), and Interstellar Glory (星际荣耀). The \$2,000/kg number follows Ref. [74] by assuming \$27 mn build cost for an (upper-stage) Starship on one-way non-reusable trip to Mars, and \$10 mn/launch for reusable Starship booster launches to Earth orbit (ten refueling flights to allow Trans-Mars Injection, eleven flights total), and then adds 46% profit margin for flights to Mars. The \$100/kg number follows Ref. [74] by assuming \$10 mn/launch for reusable Starship launches to Earth orbit but does not allow for a profit margin, based on the assumption that there will be more vigorous price competition for access to Earth orbit than for access to Mars.

(shipping solar panels from Earth), and a less expensive, unproven option (making solar-power infrastructure on Mars) [75][5].

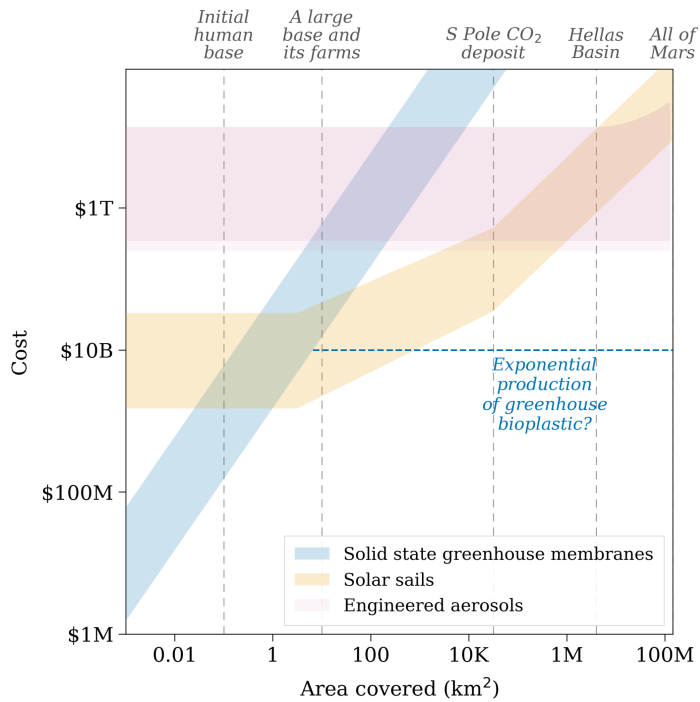

Fig. 5. A current best guess of how costs for warming Mars's surface might vary as a function of area warmed and warming approach taken (§3). We are comparing a best-guess as to how one low-Technology Readiness Level (TRL) system will scale up against a best-guess of how another low-TRL system will scale up, so rank orderings of different approaches may change as more resources are invested in technoeconomic analysis. See Appendix E and [156] for details. The dashed line shows one R&D pathway that might radically reduce costs (other cost-reduction paths are discussed in §6). The threshold used for warming is 70 sols with diurnal average temperature above the freezing point, corresponding to an energy input of ~100 W/m$^2$.

| Impact on program viability if this uncertainty resolves negatively | | | | | | | |
|---|---|---|---|---|---|---|---|
| | Existential | Detailed TEA  Climate feedback analysis | Mass production of Mars-warming aerosol | Hosted payload process experiments on commercial Mars payload services | Decision made for a human base on Mars | Integrated ecosystem tests | Create ice-covered lake |
| | Major | End-point modeling  Biocompatibility research | Solar sail space deployment test (cislunar)  Earth in situ resource utilization testing | Solar sail flight test (interplanetary) | | | Pilot factory  Support human bases w/orbiting reflectors |
| | Minor | Mars chamber testing | | Climate monitoring from orbit | | Climate monitoring network | |
| | | <$300k | $3M | $30M | | $300M | $3B |
| | | Approximate cost to retire this uncertainty | | | | | |

Table 1. Consolidated roadmap tasks, expressed using a decision relevance / uncertainty

---

[5] Radioisotope thermoelectric generators are used by NASA rovers and radioisotope heater units are used by European and NASA rovers, but neither scales to powering human bases on Mars. Fission reactors are baselined by NASA for human missions to Mars, and space fission reactors have been flight tested and ground tested (KRUSTY) and new versions are also under development. It is possible that small Mars-surface fission reactors may in the future become a cost-competitive option relative to shipping solar panels from Earth.

*research prioritization / cost framework. TEA = techno-economic analysis.*

To minimize expense, on-Mars infrastructure would need to run robotically with minimal human intervention, favoring simple methods with few consumables. Meeting all these constraints simultaneously is unlikely in early systems, which could drive up costs by a factor of several. The need for high spacecraft-level reliability, and the return-payload requirements for human missions do not apply to warming systems. Our timelines assume commercial Mars payload services begin in the 2030-2031 launch window. However, it is possible that some payloads, such as the Italian Space Agency's Mars plant growth experiment, will fly sooner.

## 3. Details of candidate Mars-warming approaches

### 3.1. Warming Mars with surface membranes

*"[T]o explore and settle the inner Solar System, we must develop biospheres of smaller size, and learn how to build and maintain them"* - National Commission on Space (1986)

#### 3.1.1. How it might work

Local warming would be cheaper than global warming. Solid-state greenhouse materials transmit sunlight and trap heat via low thermal conductivity and high infrared opacity. Aerogel achieves >60 K warming at ~3 cm thickness [28] and is already used for thermal control on Mars rovers. Non-aerogel options include wire meshes, transparent conducting oxides, and optically engineered woven canopies (Fig. 6), using multiple impermeable layers separated by Mars' thin air [76]. Water ice is also a candidate material due to its greenhouse warming effect and deep space radiation shielding properties [77].

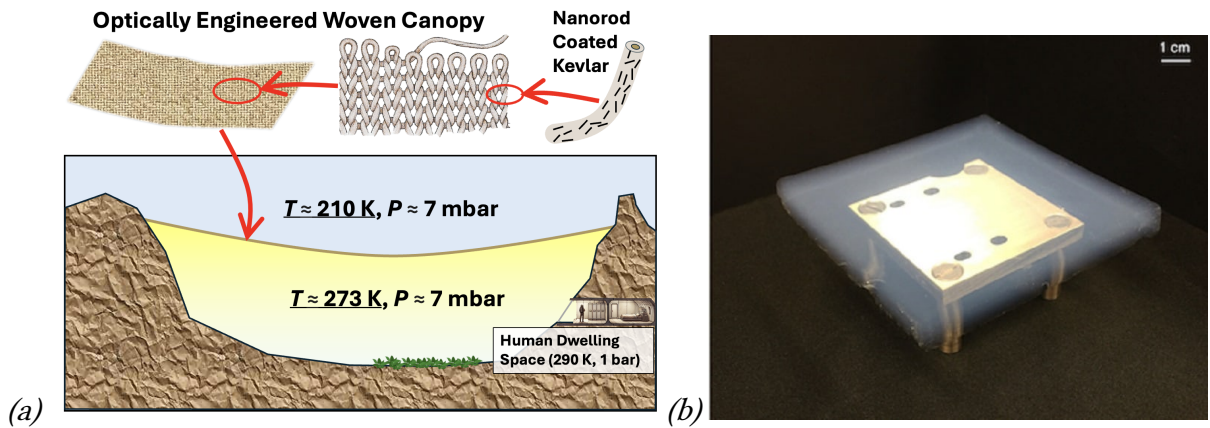

*Fig. 6. Local warming approaches. Separation of microbial habitat and human habitat allows large areas to be made habitable for relatively low cost. (a) Nanorod-coated membranes [H. Mohseni and E. Kite]. (b) Example of Mars-warming material [28].*

A near-term application could be moisture farming: warmed ice (Fig. 7) sublimates abundant vapor that, captured and condensed, provides water for life support, irrigation, fuel, and polymers. If validated, this approach could complement or eventually replace Rodriguez well technology [78-79]. It might also serve as a component of a search-for-life experiment.

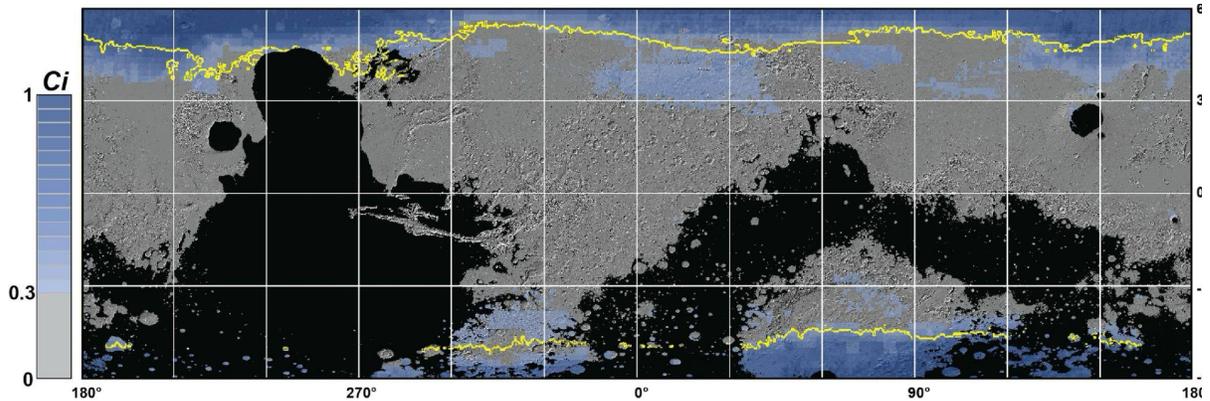
*Fig. 7. Water ice map, from [80]: Relative confidence for presence of shallow (<1-m deep) water ice at "with equatorward extent of theoretical 3-m ice stability (yellow lines) [...] black pixels mask elevations >1 km."*

### 3.1.2. Research needs

*Saving water.* Mars has much less confirmed water than does Earth: about $5 \times 10^6$ km³. Initial climate model calculations suggest water ice distribution does not change much under global warming, but this depends on assumptions [36]. For local warming, escaped $H_2O$ migrates to cold traps and is irreversibly lost from the local environment. Barriers would have to be designed and tested—impermeable membranes, condensation traps (cold fingers), or selectively permeable membranes that vent $CO_2$ but retain $H_2O$.

*Scaling.* Shipping large areas of membranes from Earth is costly (aerogel at 3 kg/m² × 1 km² of Mars surface = $3 \times 10^6$ kg, ~$6 bn for shipping excluding manufacturing cost). It is possible that non-aerogel options, such as ice or polymer-based approaches, could reduce this cost. Another alternative is making aerogel on Mars via sol-gel methods, at an estimated energy cost of 30 MJ/kg (R. Wordsworth, via email). As ~3-cm thickness is needed and aerogel density is ~100 kg/m³, 1 MW of solar panels coupled to an aerogel plant might enable a warmed Mars surface at a rate of 0.3 km²/yr. This requires supercritical drying, or surface modification for ambient pressure drying—both are an unsolved engineering problem for Mars. Biological aerogel manufacture is unproven but could allow exponential production; candidates include patterned bioplastic [81] and cellulose aerogel [82]. Several elements of the bioplastic habitat growth loop (Fig. 19) have been demonstrated in the lab [29], but the full cycle—growth, extraction, fabrication, and replication—has not. The full loop requires growing organisms under Mars conditions, extracting polymer at sufficient purity, fabricating new habitats, and repeating.

### 3.1.3. Research sequencing

*On-Earth testing (Year 0 to Year 5):*
To rapidly reduce uncertainty about the feasibility of local Mars warming, the priorities could be designing a moisture-farming mission and assessing scalability. Ranked by approximate cost (lowest to highest), this would consist of:

- Process design to assess feasibility of on-Mars aerogel production. Gating criterion: 100-ton payload gives break-even relative to direct shipment within 5 years after production begins.
- Lab test ice melting using membranes. Gating criterion: in regolith-simulant-over-ice Mars-chamber tests, <10% unfrozen ice lost to sublimation, >90% collectable meltwater.
- Earth-based optical and thermal measurements of candidate bio-polyesters to downselect membrane materials. Gating criterion: ≥70–80% average visible transmissivity across the photosynthetically active radiation range (400–700 nm), combined with infrared attenuation at Mars-warming-relevant wavelengths. Determine trade-off between selectivity (for wavelength spectrum) and density, gas-permeability, elastic modulus/toughness, etc. Co-polymers of short-chain aliphatic (e.g. glycolic) with arylaliphatic (e.g., mandelic or phenyllactic) or aromatic (e.g. phloretic or benzoic) hydroxy acids might be candidates for bio-producible materials to manufacture membranes that retain heat while transmitting visible light.
- Seek lighter solid-state greenhouse materials that could warm Mars by 10s of K. Gating criterion: <1 kg/m$^2$ in Mars-chamber tests, 3× better than the current state-of-the-art.
- Mars-chamber tests of bioplastic habitats. Gating criterion: show all steps needed for 3D-printer-assisted exponential growth of bioplastic habitats with realistic temperatures and day–night cycles.
- Moisture farming de-risking: lightweight barriers trapping $H_2O$ vapor.
- Developing a Mars test mission concept.

The highest scientific risk is achieving sufficient bioplastic transparency and yield (>16% of biomass [29]) for exponential production of habitats. If achieved in the lab, this would still have to be shown to scale—leading possibilities are cellulose acetate or a PHA/PLA (polyhydroxyalkanoate/polylactic acid) copolymer[6]. If on-Mars production proves infeasible and membranes do not scale to large fractions of planet surface area, membranes could still support initial bases.

*In-space testing (Year 5 to Year 10):*
- Flight demonstrate greenhouse-membrane materials for warming Mars, for example at the Moon. Gating criterion: Warm soil by >35 K.
- Flight demonstrate moisture farming at >10 m$^2$ scale.
- Flight demonstrate biomaterial greenhouse.

These mission concepts are starting points; mission design will evolve as research matures. Analogous NASA lunar surface experiments as secondary payloads cost $4-$50 mn; for Mars surface experiments, we estimate $30 mn with uncertainty of a factor of two. Priority order could be set by the extent to which benchmarks are exceeded during on-Earth testing.

---

[6] For the latter, ester-bond density and backbone crystallinity jointly govern infrared opacity: natural poly(3-hydroxybutyrate) is too crystalline and mechanically brittle for membrane applications, but copolymer strategies incorporating straight-chain ω-hydroxyacid comonomers could reduce crystallinity and improve flexibility while simultaneously increasing ester-bond density and infrared attenuation at Mars-warming-relevant wavelengths.

### 3.1.4. What could be enabled by research

If research succeeds, activities that could become feasible include harvesting 500 m$^3$ of liquid water (enough to fuel a human-capable lander); testing the capability for exponential production of habitats via 3D-printer-assisted habitat exponential growth; and creating a small ice-covered lake on Mars (Fig. 8). A key decision point would then be go/no-go for inoculation with Earth-derived microbes in semi–closed environments.

Although most of these steps could be carried out under existing COSPAR Category IVa bioburden recommendations, the last step would likely invoke extensive discussion at COSPAR and elsewhere about whether local inoculation of semi-closed environments offered adequate protection for other places on the planet. It is also possible that Mars planetary protection categories will continue to evolve in the context of human exploration, as has been the case for the Moon.

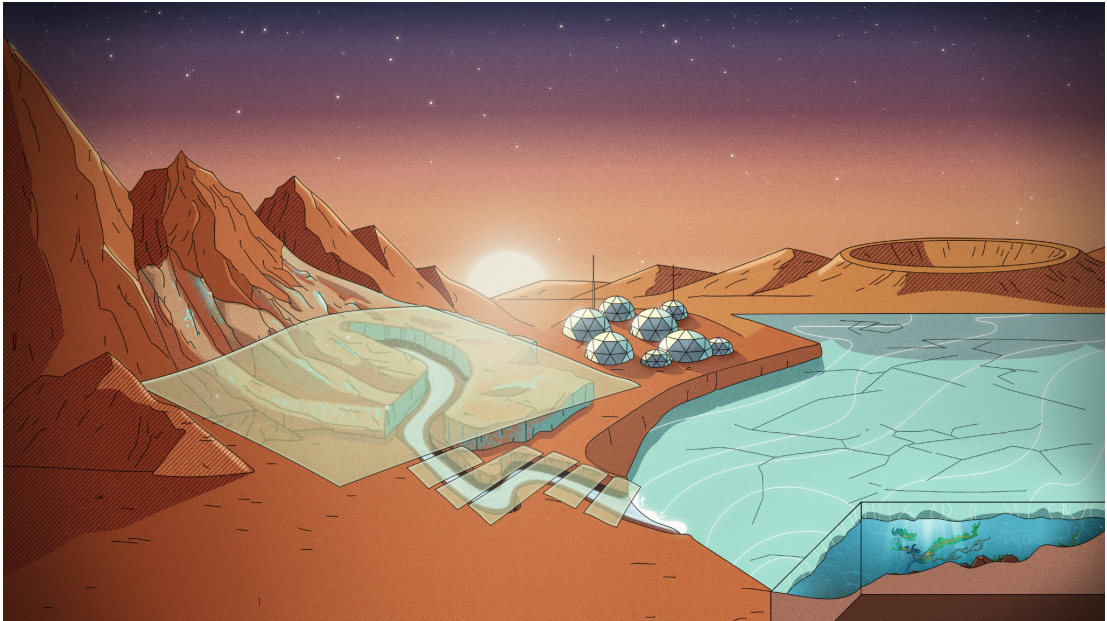

*Fig. 8. Artist's impression (V. Socianu) of an ice-covered lake on Mars fed by meltwater runoff from a debris-covered glacier. Solid-state greenhouse membrane on the glacier causes sublimation and melting. Debris-covered glaciers are common on Mars—three of the top seven candidate SpaceX landing sites adjoin one [83]—and are >80% water ice. Confined by soil ridges, meltwater accumulates as an ice-covered lake sustained by continued melt [84]. The ice cover could be coated with low-vapor-pressure liquids to slow sublimation. Conditions would resemble those of permanently iced-over Antarctic lakes [85]. Photosynthesis in the lake might support ecological studies and produce biomass and oxygen.*

## 3.2. Illuminating Mars with orbiting reflectors

In 1993, Syromyatnikov and colleagues used a cargo spacecraft to unfurl a >300 m² mirror, sweeping a 4-km wide beam across Europe brighter than the full Moon. In the context of Mars exploration, orbiting reflectors might provide illumination, sunlight power, and warmth to human bases (Fig. 9). This is especially beneficial during the Mars winter, when thick $CO_2$ ice build-up can permanently damage structures[7]. Cost-effectiveness depends on reflector mass, lifetime, and manufacturing assumptions that remain untested.

### 3.2.1. How it might work

Orbiting reflectors can warm small surface regions [35,86-87][8]. Orbiting reflectors might be cost-competitive with solid-state greenhouse warming for a growing human base. For the first human missions to a midlatitude site, a 1 km² orbiting reflector can only provide <1 W/m² extra time-averaged power due to orbital mechanics and the finite angular size of the Sun, even for a perfectly smooth reflector, which is unlikely. Hundreds of km² would be required to double the insolation of a human base. However, for early Mars bases, materials would be manufactured on and delivered from Earth, so the ability of solar sails to fly themselves from Earth orbit to Mars propelled by solar radiation pressure (a mechanism demonstrated by JAXA's IKAROS) removes the cost of both trans-Mars injection and Mars entry, descent, and landing. Moreover, the thin atmosphere of Mars allows microclimate warming to be confined. In areas of steep terrain, Mars-surface-mounted reflectors might be cost-competitive with orbiting reflectors in creating favorable microclimates.

If a local region is warmed on Mars, $H_2O$ will be lost via sublimation and will migrate to spatially distant cold regions unless the $H_2O$ is contained. This can be mitigated by warming locations with abundant water ice, such as buried glaciers in flat terrain that outcrop on scarps as ice cliffs.

From a sun-synchronous polar orbit with altitude ~750km, 1 km² of reflector surface can provide a Mars-year-averaged ≈0.6 MW of power to a 1-km-radius region of interest (ROI) located at 40°N. (This assumes that when the ROI is visible to the sail, the sail points to optimally reflect light onto the ROI; otherwise, the sail orients to station-keep and correct orbital perturbations.) Doubling (in an average-power sense) the amount of sunlight reaching such an ROI would require ~750 km² of reflector surface in orbit and would cost ~$3.5 bn assuming $100/kg launch cost, $100/kg manufacturing cost, 20 g/m² spacecraft areal density, and $0.5 bn research and development cost. Cost falls to ~$2 bn for 10 g/m² mass. The reflectors would resemble a slowly twinkling ring around Mars.

---

[7] https://www.jpl.nasa.gov/news/phoenix-mars-lander-is-silent-new-image-shows-damage/

[8] An illumination demonstrator, Reflect Orbital's EARENDIL-1, is scheduled for 2026 launch. Other orbital approaches do not work: IR-resonant rods in Mars orbit have short collision timescales for the needed Mars-warming number densities. IR-resonant wire meshes in Mars orbit are short-lived due to radiation pressure deflection of the orbit.

Appendix D discusses the polar warming application (raising atmospheric pressure by sublimating $CO_2$ ice [13]), which is separate from (and more challenging than) warming a human base.

We assume a sun-synchronous polar orbit, which provides continuous Sun visibility and frequent overflights of all points on Mars' surface.

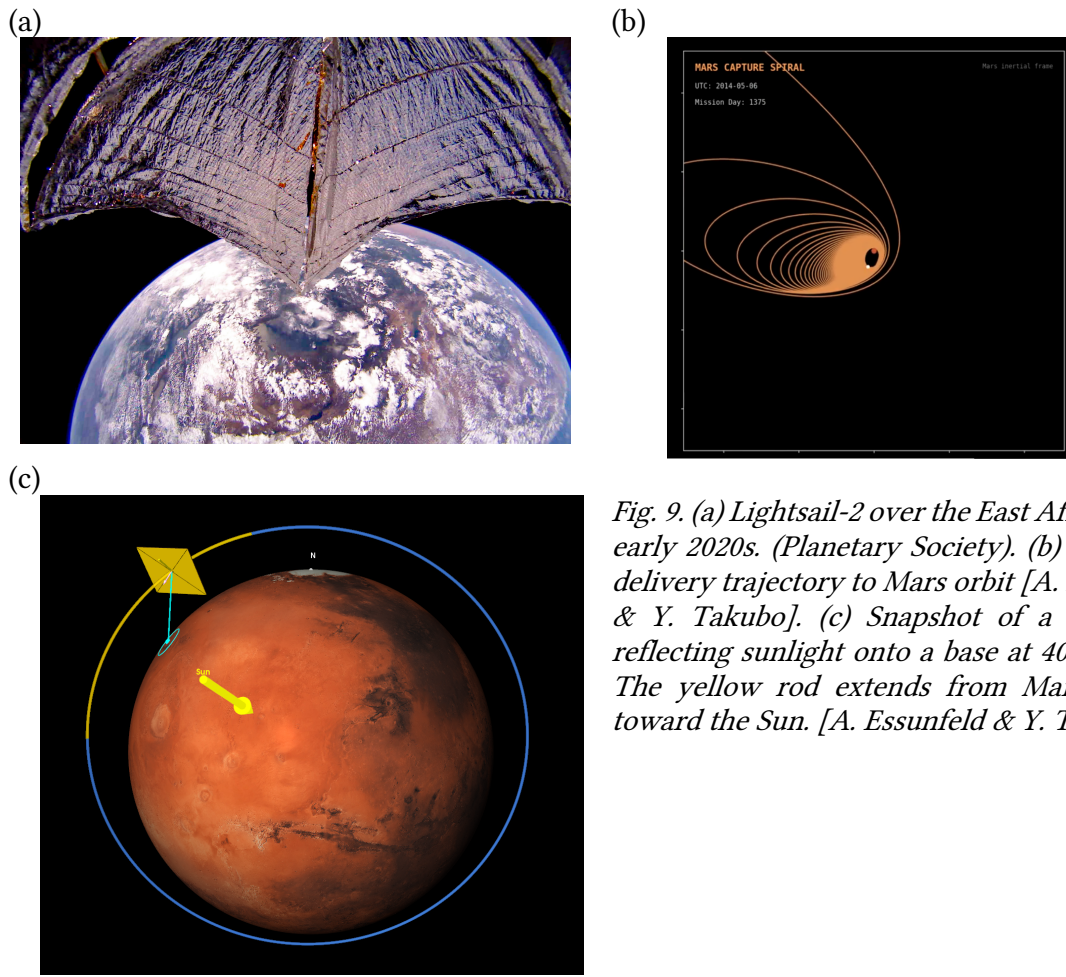

*Fig. 9. (a) Lightsail-2 over the East African Rift, early 2020s. (Planetary Society). (b) Solar sail delivery trajectory to Mars orbit [A. Essunfeld & Y. Takubo]. (c) Snapshot of a solar sail reflecting sunlight onto a base at 40°N 200°E. The yellow rod extends from Mars' center toward the Sun. [A. Essunfeld & Y. Takubo].*

### 3.2.2. Research needs

To be feasible for warming Mars, sailcraft would need to achieve ≤20 g/m², which is 3× lower than the as-constructed state-of-the-art (Fig. 10) [86,88,89]. Further reduction would require in-space testing. The sail component of a sailcraft may be low-mass: as-built sails mass <4 g/m², and candidate membrane materials have mass <0.5 g/m² [90-92]. However, a sailcraft that serves as a reflector would also need to:

- Communicate with Earth, aiding Doppler tracking and thus navigation.
- Direct reflected light efficiently to the targeted region without sail ripples.
- Control orientation, e.g., with reaction wheels or by modulating sail reflectivity (as demonstrated by IKAROS).

Bigger sails or formation fliers might amortize spacecraft-function mass [86] (allowing low g/m²), but would require complex deployment. As launch fairing sizes increase (e.g., 8.7 m for New Glenn 9×4), rigid or pre-deployed disks are worth investigating.

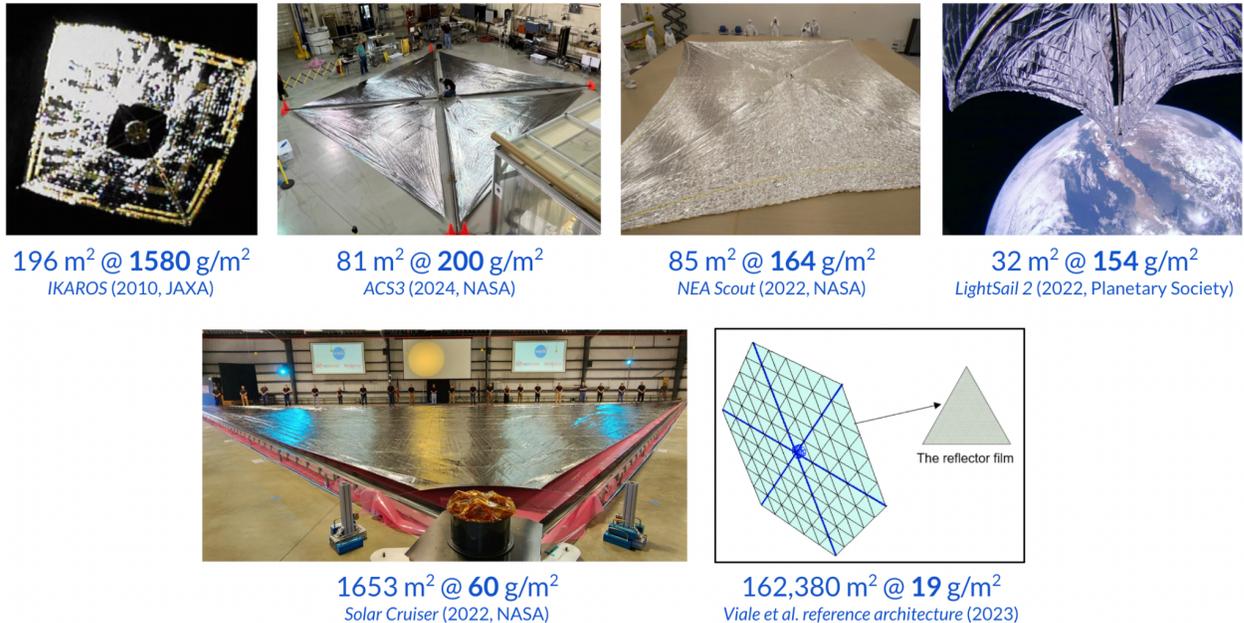

196 m² @ **1580** g/m²
*IKAROS (2010, JAXA)*

81 m² @ **200** g/m²
*ACS3 (2024, NASA)*

85 m² @ **164** g/m²
*NEA Scout (2022, NASA)*

32 m² @ **154** g/m²
*LightSail 2 (2022, Planetary Society)*

1653 m² @ **60** g/m²
*Solar Cruiser (2022, NASA)*

162,380 m² @ **19** g/m²
*Viale et al. reference architecture (2023)*

*Fig. 10. Examples of flown, constructed, and concept solar sails.*

The trade-off among sailcraft number, size, complexity, and cost has not been studied, making this a priority. Ground tests are of limited value[9], so de-risking would require flight tests. One plausible development path is discussed below:

- A minimum viable at-Mars orbiting reflector illuminates a surface asset. To do this, a 5 m-diameter orbiting reflector would suffice, which is CubeSat-deployable. (NASA has demonstrated Mars flyby CubeSats with MarCO; ESA's LightShip is designed to release CubeSats into Mars orbit).
- The next step would be sailcraft flight from Earth orbit to sun-synchronous Mars orbit. Either or both might piggyback off already-planned pathfinders such as the 20 g/m² Sundiver CubeSat concept [88,89].
- The next leap would be to provide supplemental sunlight to a human base. Doubling sunlight (a useful yardstick) would require ~20 km² of reflectors in view at all times. Taking into account duty cycle, ~750 km² of reflectors would be required, at a total launch cost (ignoring manufacturing cost) for 10 g/m² of approximately $2 bn. This greatly warms a city-sized area. (Mars' thin atmosphere cannot effectively convect away the energy from warming). 750 km² of reflectors at 1-10 g/m² would have a combined mass of 2 to 18 times that of the International Space Station. Mass production at this scale unlocks more ambitious applications (Appendix D).

---

[9] The Fallturm Bremen allows 9 seconds of near-vacuum zero-G conditions, but only for <0.8 m diameter payloads.

### 3.2.3. Research sequencing

For solar sails, a bottleneck is that ground testing does not adequately de-risk in-space deployment of sails (do-or-die megascale reverse origami) [92]. Solar sails have encountered difficulties at deployment, and late in development. Simulation and ground test are necessary but insufficient. The overall goal is to determine the cost for making >$10^3$ km$^2$ combined area of such sailcraft. This will require orbital tests to find the most mass-efficient sailcraft that is robustly flyable for long-duration Mars warming. It is also of interest to determine if rigid disk or pre-deployed sailcraft can side-step the deployment issue.

A line of flight opportunities for solar sails would benefit both Mars warming research and other applications [86,88]. Orbits at Earth would need to be >800 km high to minimize drag.

*On-Earth research (Year 0 to Year 3), listed from least to most costly:*
- Assess how Mars-orbiting sails should be disposed of at end-of-life (raise versus lower orbit). Go/no-go criterion: No practical impact on planetary protection nor collision risk.
- Analyze weather feedbacks of local warming, including the effect of clouds, albedo changes, latent heat, and quantifying reversibility. Go/no-go criterion: does doubling the sunlight received by a human base in 15 years cost <$0.5 bn/yr when climate feedbacks are included?
- Co-design of an orbit and sailcraft that allows a stable, useful orbit around Mars with slew control sufficient to warm selected patches of Mars. This would include determining whether to send a small number of large spacecraft, or many small spacecraft. Go/no-go criterion: If a sailcraft design cannot fly to Mars for ≤20 g/m$^2$ total areal density, sail warming is unlikely to be affordable for Mars.
- A pathfinder-mission detailed design effort (e.g., [88]) might include: (i) design of demo reflectivity-control-device-based attitude control at relevant angular momentum scales. Alternatively: demo larger sail with "bulky" attitude control (i.e., not just modulating solar sail reflectivity). (ii) Design of demo spiral-up-from-Earth, navigate-to-Mars, spiral-down-at-Mars. (iii) Design of demo precise & accurate targeting of reflected beam. (iv) Design of demo deployment (if deploying).

An analogous NASA-funded interplanetary solar sail design effort was done in two years for ~$2 mn.

*Earth-orbit research (Year 3 to Year 5):*
- Flight test in cislunar space of a pathfinder mission spacecraft. Gating criterion: demonstrate controlled flight with 1 km/sec of total photon-induced delta-V, breaking the JAXA-held record. For a 20 g/m$^2$ sailcraft, this defines an endurance test of >1 month.

*At-Mars testing (Year 5 to Year 10):*

- Pathfinder mission reflects sunlight to Mars' surface, appearing brighter than the brightest star in Mars' night sky, in order to confirm end-to-end pointing control.
- Launch of Earth-to-Mars solar sail flight ending in Mars sun-synchronous orbit with 100 km precision in target orbit.

The two goals could be achieved in a single mission with one spacecraft, but this may lead to a complex architecture. The second goal could also be achieved by a CubeSat deployed from Mars orbit (by, e.g., the European Space Agency's LightShip). Although the Planetary Society's successful philanthropically-funded LightSail project cost $7 mn, this is very likely too small for the missions discussed here. Ambitious solar sail missions have historically had budgets in the low tens of millions of dollars (when government-managed). This could be suitable for a competition or government-funded challenge.

A future, large-scale test could support a human base by doubling the sunlight that reaches that base, for ~750 km$^2$ of solar sails. More ambitious applications are discussed in Appendix D.

These would all be COSPAR Category III missions (under Mars' current categorization).

## 3.3. Warming Mars with engineered aerosols

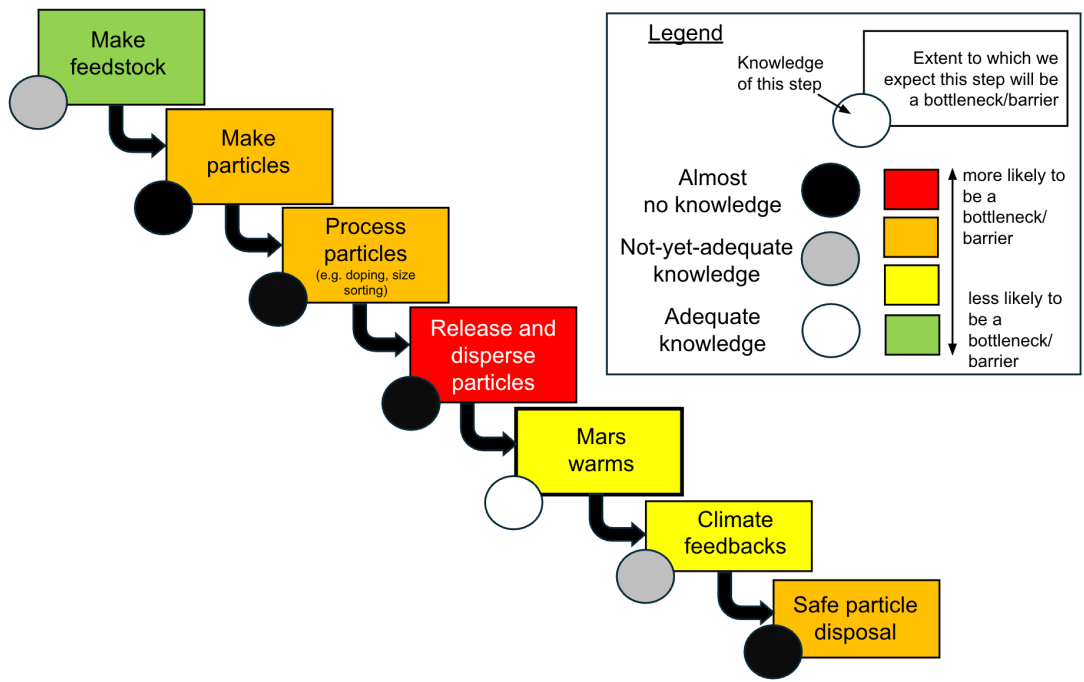

*Fig. 11. Challenges of aerosol warming for Mars: March 2026 summary of knowledge gaps.*

### 3.3.1. How it might work

A few million tonnes of atmospheric aerosol can significantly alter a planet's energy balance [64]. In atmospheric science, 'aerosol' refers to any suspended particle; the particles considered here are purpose-designed, with controlled composition and size. This makes aerosols appear to have potential for a relatively inexpensive climate process experiment (such as warming a region or the globe to study climate feedbacks). However, to do this particles would need to warm the planet (right size and shape), stay dispersed (resist agglomeration), have an effective lifetime long enough to have the desired climate effect despite surface deposition, and degrade without harming health or the environment. Experimental data on each of these steps are sparse (Fig. 11).

Particles could be made from Mars' air, or from its soil and rocks. Each approach has its own challenges and some problems are common to all methods (Table 2).

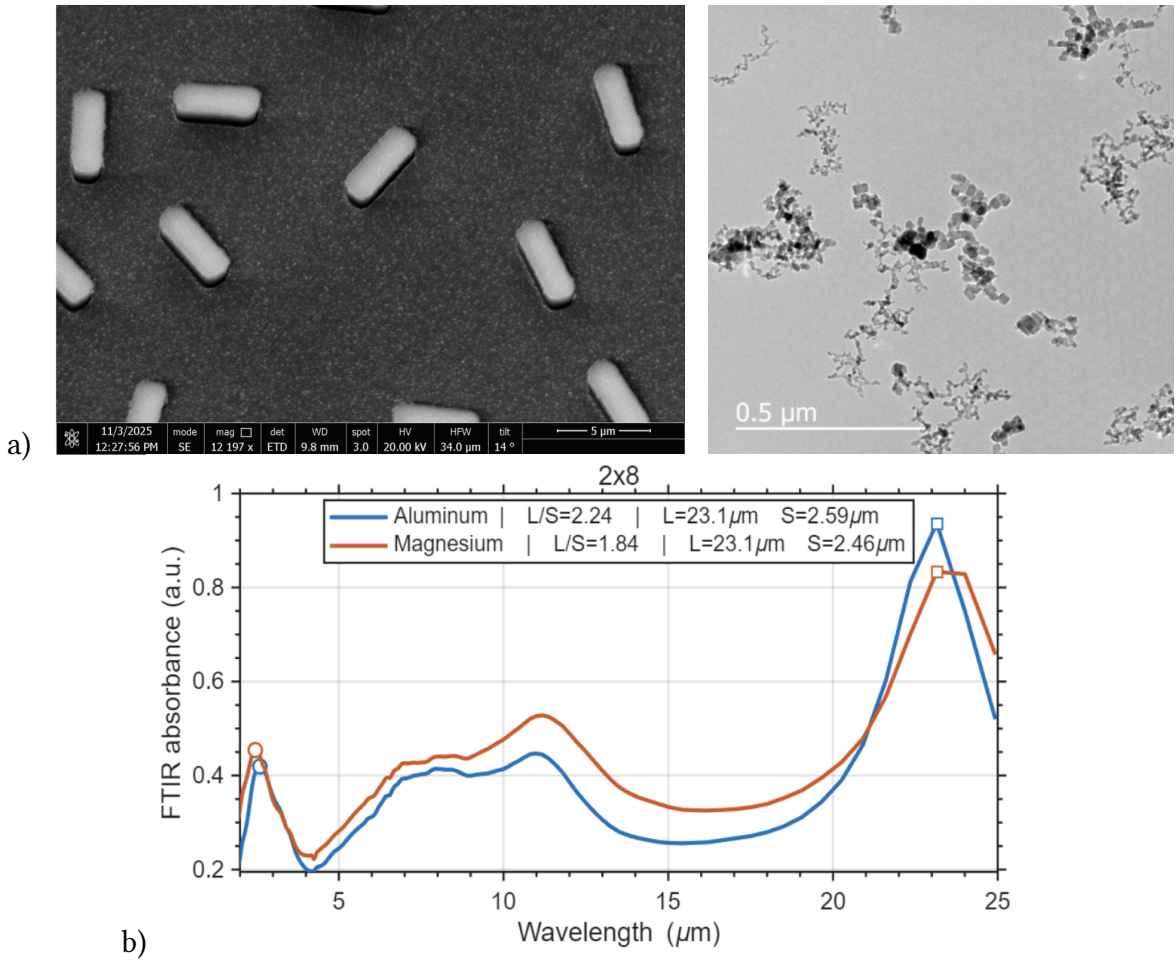

Fig. 12. (a) Test batches of small particles intended to demonstrate Mars-warming properties. Left: Randomly oriented magnesium nanoribbons on negative-lift-off pillars. (H. Mohseni/A. Bamba/S. Ansari, Northwestern University) Right: Mg-aggregates made using aerosol-phase production processes (A. Boies/C. Jourdain, Stanford University). (b) Fourier transform infrared (FTIR) extinction spectra for 2 μm-wide, 8 μm-long nanoribbons. L/S is the ratio between the peak absorbances (circled) at long (L) and short (S) wavelengths. A. Bamba & H. Mohseni, Northwestern University.

| *Key open questions for using atmospheric carbon as feedstock.* Mars-warming net radiative effect of carbon-based aerosol has not been demonstrated in the laboratory. | *Key open questions for using metals from soil and rocks as feedstock.* Safe particle degradation. Complex production pathway. | *Natural minerals (e.g., salts, silicates) as feedstock.* Large column mass required → clumping concern. Lack of data for Mars-abundant salts other than halite and gypsum. |
| --- | --- | --- |
| *Shared open questions (and how to answer them).* Show that effective particle lifetime can reach ⅓ years: (1) Release and dispersal in Mars-relevant environmental conditions (Mars-chamber test). (2) Dust feedback (modeling). (3) Charge state of atmosphere (on-Mars measurement). (4) Dry deposition rate (Earth and Mars tests). (5) Photophoresis (modeling). (6) Re-lofting (Mars-chamber test). (7) Contact angle parameter (Mars-chamber test). | | |

Table 2. Shared and specific challenges for different particle production pathways.

**Using Mars' air as feedstock (Fig. 13a).** The Mars-warming net radiative effect of carbon-based aerosol has not been demonstrated in the laboratory, although graphene disks and ribbons (size 0.25-1 µm) resonate strongly with Mars infrared radiation in experiments [20,93]. Models predict powerful greenhouse warming, although lab validation is needed for the net radiative effect of carbon-based aerosol. Mars air might be used as feedstock to make carbon particles[10] (Fig. 14, Appendix A). Mars air is 95% $CO_2$; a third of atmospheric atoms are carbon. Carbon-based aerosols affect Earth's energy balance (e.g., via albedo reduction of ice surfaces [94]), although these specific carbon-based aerosols absorb sunlight that then cannot drive photosynthesis—making them unsuitable for enabling biospheres. Carbon is engineered into many forms including nanotubes, fullerenes, and graphene. MOXIE [95] demonstrated in situ $CO_2$ electrolysis to CO and $O_2$ (Fig. 14). Exothermic disproportionation (CO + CO → $CO_2$ + C) yields carbon, including graphene [96]. Assuming N-doping to +0.6 eV [20] and 1-year particle lifetime $\tau$, the power requirement for doubling the strength of Mars' greenhouse effect (specifically, warming the planet by ~4 K) is:

$$P = E_{kg} \, A_{Mars} \, m_c \, / \, \tau \approx 500 \text{ MW}$$

where $E_{kg}$ is the energy-per-kg of making Mars-warming particles (~80 MJ/kg), $A_{Mars}$ is Mars surface area, and $m_c$ (1.3 mg/m²) is the column mass in carbon-based particles needed to warm Mars by ~4 K [20]. This is less than the total power that humans plan to place in space in the near future—but still challenging given <1 kW total power at Mars' surface as of 2026. The same model outputs that ~5 GW would be needed to warm Mars by 35 K (i.e., sub-linear scaling—each additional K of warming would require more power), but this depends on the amplitude of climate feedbacks that a climate process experiment would be needed to discern.

---

[10] Where measured, Mars soils and rocks have ½ kg/m³ of organic carbon, which could be an alternative source of reduced C (CO or $CH_4$) [17].

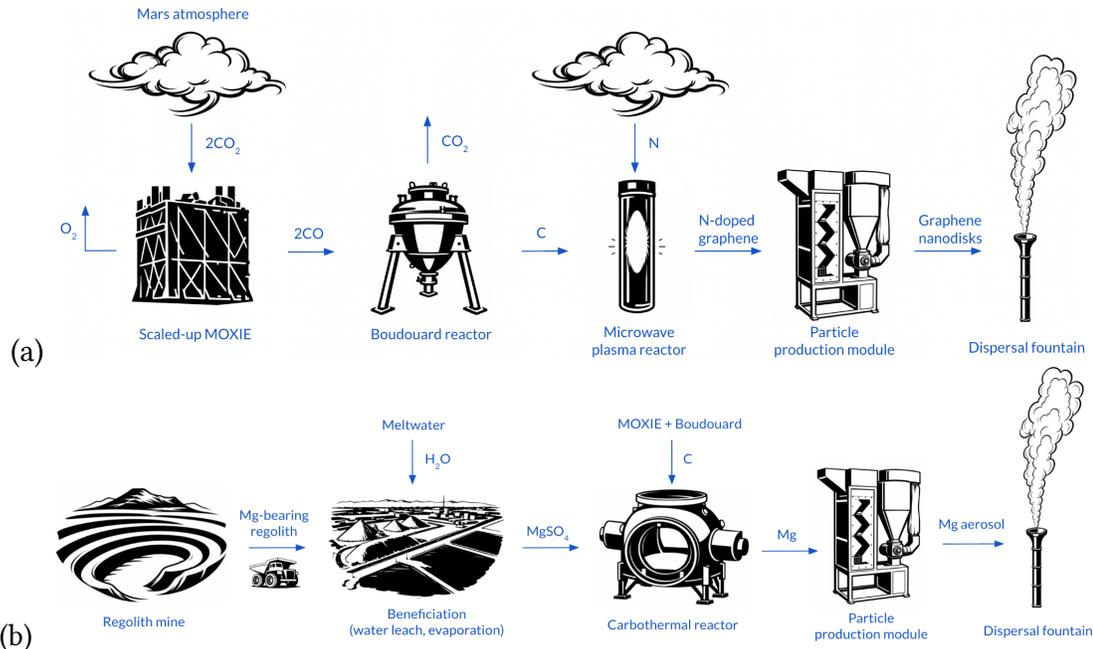

Fig. 13. Concept of operations sketches for warming Mars using (a) air as feedstock, or (b) using soil and rocks as feedstock.

Using commercial solar panels (73 g/m$^2$ - 145 g/m$^2$ is advertised by Starpath, as-flown solar panel systems mass 2-3 kg/m$^2$), the mass for 500 MW of power at 40° N is 5000 tonnes. This could be lowered by on-Mars panel production, or concentrated solar power with Mars-made reflecting elements. Overall, power system mass is unlikely to be the binding constraint.

Particle lifetime is the biggest uncertainty. Required effective lifetime is ≥4 months; no Mars-relevant measurement exists. On Earth, tropospheric aerosols last ~1 week, and stratospheric aerosols last ~1 year. Martian global dust storms decay in ~0.25 years. Mass-efficient warming would require numerous small particles, which are less quick to settle, but more likely to clump. Uncertainty could be reduced by theoretical calculations fueled by laboratory measurements of key aerosol parameters (Appendix A).

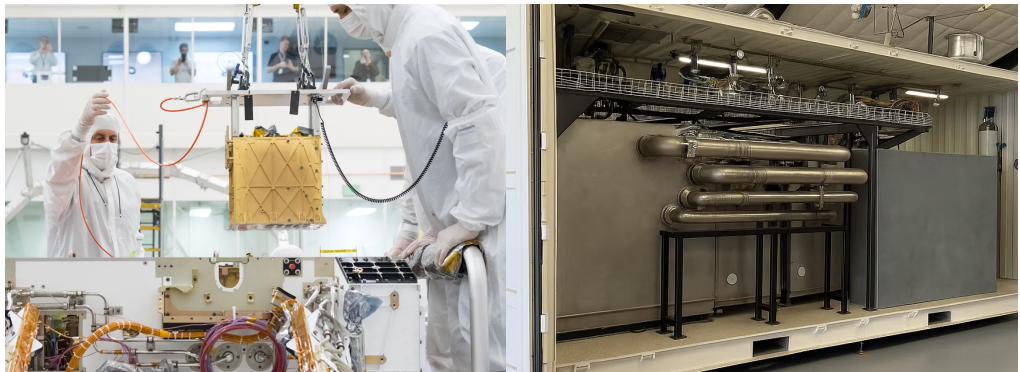

Fig. 14. (a) MOXIE instrument (JPL). (b) Carbon energy storage system (from noon.energy).

**Using Mars soil and rocks to get metals to use as feedstock (Fig. 13b).** For metals, Mars-warming radiative properties have been experimentally validated. Conductive particles (ribbons, rings, rods, disks) can resonate with Mars-warming-relevant infrared wavelengths [97-98]. Test batches were fabricated (Fig. 12a), and Fourier Transform Infra-Red spectra (H. Mohseni and A. Bamba, Northwestern University) validated the simulations of particle radiative properties (Fig. 12b).

In climate models, 2 mg per m$^2$ of atmospheric column of conductive particles doubles the strength of Mars' greenhouse effect and warms Mars by ~4 K [20]. For 1 year particle lifetime, the amount of metal needed to warm Mars by ~4 K is 2 L/s and for >35 K is 60 L/s (for aluminum; simulations indicate similar volume fluxes for Fe, and experiments indicate similar volume fluxes for Mg—all three elements are abundant in Mars' soil and rocks) [20,24,99]. For further discussion of metal and particle production pathways, see Appendix A. One potential source of metals is $MgSO_4$ salts [61], which are abundant on Mars, especially in salty-rock mountains (~20 wt% [100]). Estimated salty-rock mass is ~$10^{18}$ kg [61,101]. Highly soluble, $MgSO_4$ salts can be leach-separated. Leaching requires $H_2O$ from hydrated minerals or ice; the $H_2O$ could be recycled [102].

Clumping lifetime is longer for metals than graphene (fewer particles needed) but is a major unknown given natural-dust interactions. Clumping might be mitigated by anti-stick coatings. Simulations show clumps of ~10 nanorods remain radiatively effective (Fig. 15).

### 3.3.2. Research needs

*Show that effective particle lifetime can reach ⅓ years.* Particle lifetime against agglomeration to clumps that can sediment out would depend on particle number concentration, size, and charge state (Appendix A), but models of the number of charges on Mars aerosols [103] need more at-Mars tests and electric field measurements. A re-flight of the MicroARES (lost with the *Schiaparelli* crash), or a similar sensor, is needed. Clumping might be mitigated by reducing required particle number [104,105], engineering for radiatively advantageous clumping (superscattering [106]) (Fig. 15), designing for ease of re-lofting or charging [107] (§6), or applying anti-stick coatings. If effective lifetime <⅓ year in models informed by Mars-chamber testing, this approach is not viable at global scale.

Preliminary cost central estimates are on the order of $0.1–0.6 bn/K/yr to double the strength of Mars' greenhouse effect (+4 K) (details are in Appendix A and at [156]), but with order-of-magnitude uncertainties. With the same assumptions, warming by 35 K would cost $0.4-1.3 bn/K/yr. Many assumptions underpin these cost estimates; for example, clumping is assumed not to limit lifetime, and the doping+graphene system mass may be understated.

### 3.3.3. Research sequencing

*On-Earth research (Year 0 to Year 3) and gating criteria:*

On-Earth research includes, in priority order, (1) particle downselect, (2) lab experiments on microphysical parameters and biocompatibility, (3) In situ resource utilization factory design maturation, (4) scale up batch production.

(1) Particle downselect.
- Dispersal demonstrations. Gating criterion: in simulations informed by dispersal demonstrations, >80% of particles get >1 km from release site as monomers.
- Confirm warming (Fig. 15). Go/no-go criterion: in simulations based on laboratory measurements of radiative and microphysical properties, particles warm Mars by 35 K while allowing >60 W/m² of diurnally-averaged sunlight at 50° S in summertime.
- Climate feedback analysis including dust cycle.
- End-to-end analysis of the effect of clumping on radiative properties.
- Biocompatibility and degradation assessment.

For the *air-feedstock pathway*, technical risk is front-loaded, as the Mars-warming potential of these particles is unproven. For the *soil and rock feedstock* pathway, there is also significant scientific risk (particle degradability is unconfirmed).

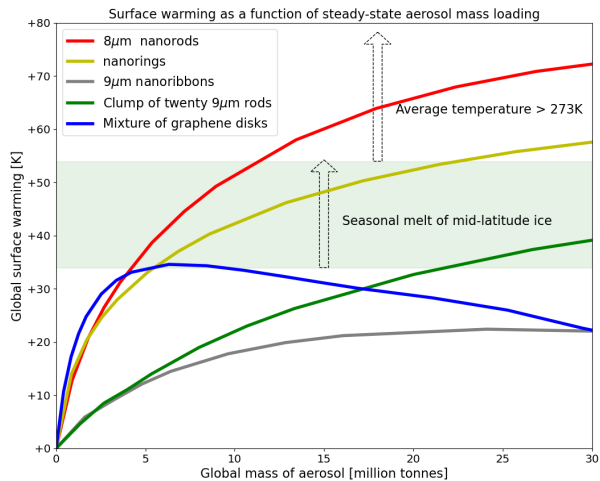

*Fig. 15. Output from 1-D climate model TerraScreen[3], showing relative strength of warming.*

Following particle downselect:

(2) Lab experiments on microphysical parameters.
- Test microphysics of particles, in priority order:
a. Measurements of dry deposition rate (removal of particles from the atmosphere onto the surface through Brownian diffusion and other processes) for aerodynamically equivalent particles on Mars analogue sites on Earth, to assess the lifetime of aerosols in the Martian atmosphere.
b. Mars-pressure wind-tunnel tests, to assess the ability for aerosols to be re-lofted back into the atmosphere once they have already been deposited onto the surface. This affects the ability for aerosol warming to be self-sustaining.

- c. Clumping measurements in the laboratory, to assess the lifetime of charged and uncharged aerosol particles subject to self-aggregation and collisions with dust.
- d. Lab contact angle tests for $H_2O$ and $CO_2$ ice, to assess the ability for engineered aerosols to nucleate cloud particles in the atmosphere, which can a) affect the ability for aerosols to warm Mars and b) reduce the lifetime of aerosols in the atmosphere.

Gating criterion: simulations using these data indicate $>\frac{1}{3}$ year effective atmospheric lifetime and that lifetime could be tuned.

- Degradation and biocompatibility testing. Gating criterion: particle deployment poses no hazard to animal or plant health.

(3) ISRU factory design maturation.
- Small scale in situ resource extraction experiments. Gating criterion: demonstrate efficient extraction of one or more relevant feedstocks from Mars-abundant ingredients.
- Technoeconomic analysis (TEA). Benchmark size and scope of the fully deployed factory [156], with cost estimates that include research-and-development costing. If the soil-and-rock feedstock pathway is chosen, implementing this pathway would require an industrial infrastructure on Mars. The minimum viable pilot factory involves multiple steps and is quite complex.

(4) Scale up batch production.
- Kg/yr-scale production. Go/no-go criterion: Verify size sorting to factor-of-4 spread and within-50% control of modal size, and (if graphene needed) verify doping.

*On-Mars research stage (Year 4 to Year 10) and gating criteria:*
- Aerosol release process experiment (in year 5). Gating criterion: Release and track >100 g of particles to >100 m altitude, with <20% of particles lost to clumping. Cost: Low tens of millions of dollars as a commercial Mars payload services secondary payload.

As for all Mars surface payloads, test missions would be COSPAR Category IVa under current categorization, unless colocated with a site of astrobiological interest.

- ISRU technology demonstrations. Necessary precursor experiments to de-risk resource extraction and particle production technologies for full scale factory deployment might also be carried out in this timeframe. Which aspects of production can be de-risked in Mars environmental chambers versus requiring flight tests is an open question.

### 3.3.4. What could be enabled by research

If the gating criteria described above are met, the following activities would become feasible. On-Earth ISRU testing would need to demonstrate a prototype pathway reducing consumables to <20% of initially deployed factory mass over 30 years, including Mars-chamber testing. Based on the number of personnel involved for a comparable ISRU project (Blue Origin's Blue Alchemist), a cost of $150 mn appears credible. Next, a pilot factory

on Mars (likely post-human-landing) would need to prove in situ element extraction and particle production. We do not know how much this would cost: this would be a heavyweight interplanetary lander payload, with few comparison points (the development of Starship cost $5B+, and the development of Blue Moon Mark 2 will cost >$7B). The next step would be to release a plume of sufficient size to confirm small-scale warming. Later, climate experiments (regional or global), if determined to be feasible and desirable, would be guided by models and reanalyses. Subscale regional or global warming experiments would create insufficient warming for habitable environments (no biological planetary protection concerns) but would modestly moderate cold temperatures at habitats, and produce measurable feedbacks to test models of warming.

### 3.4. Summary of roadmapped research

Table 3 and Figure 16 summarize the research in the roadmap.

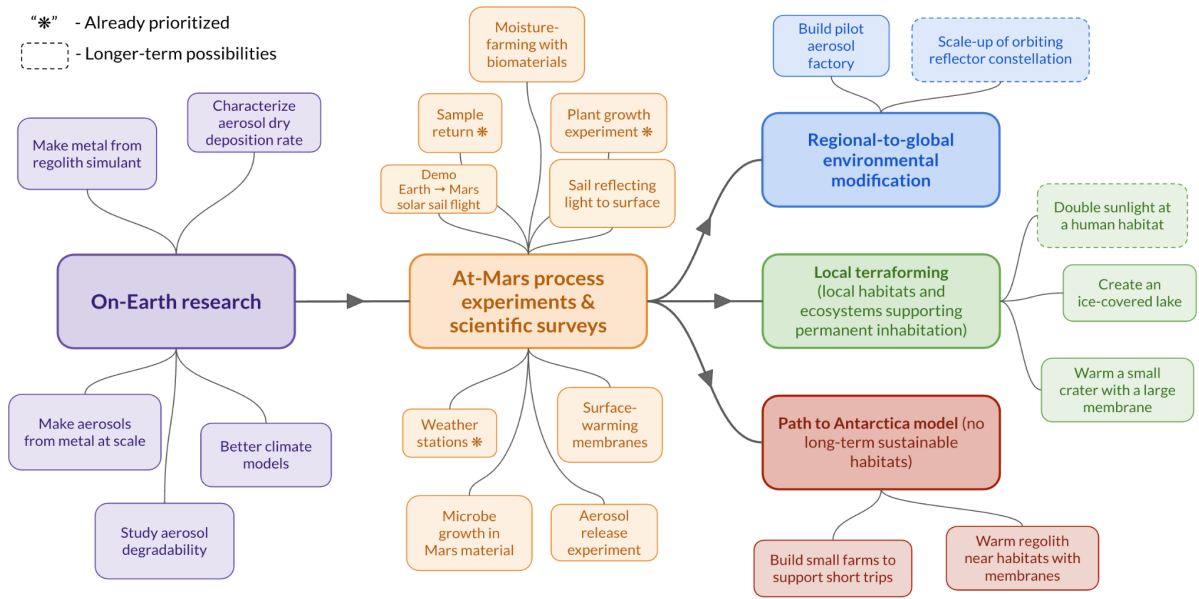

Fig. 16. Possible research projects, precursor missions, technology demonstrations, and small-scale deployments, organized by on-Earth / on-Mars research phases and different possible human-involved futures for Mars.

**On Earth (modeling, lab experiment, …):**

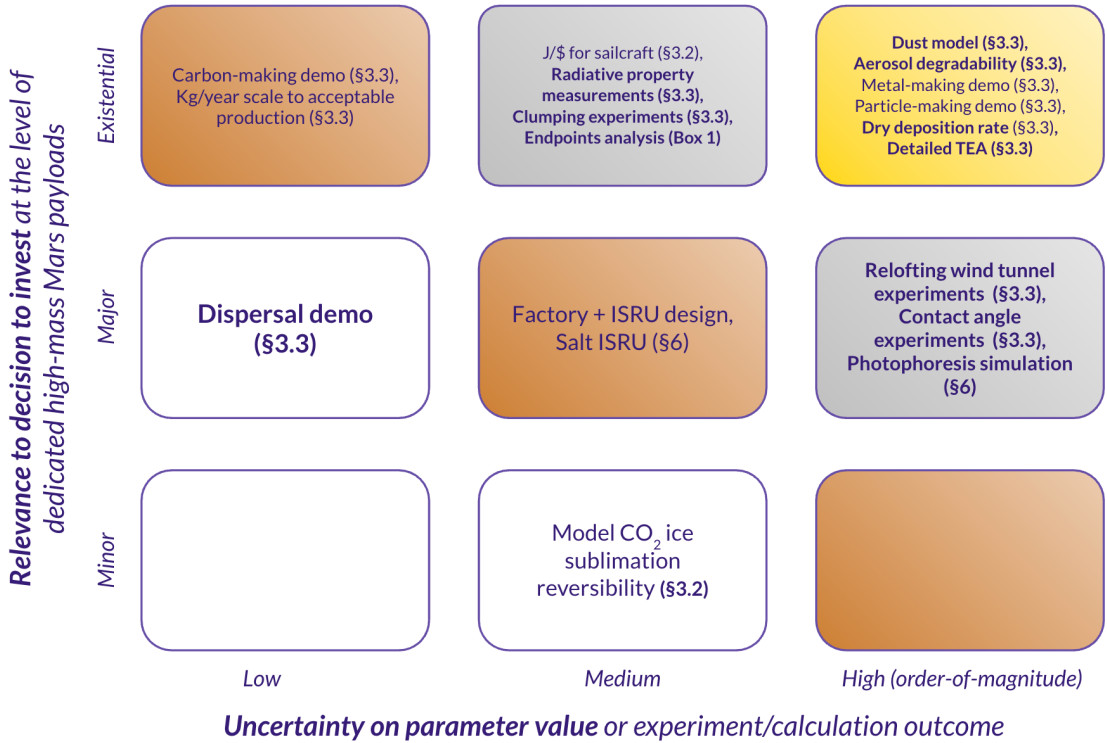

**In space (remote sensing, validation/test):**

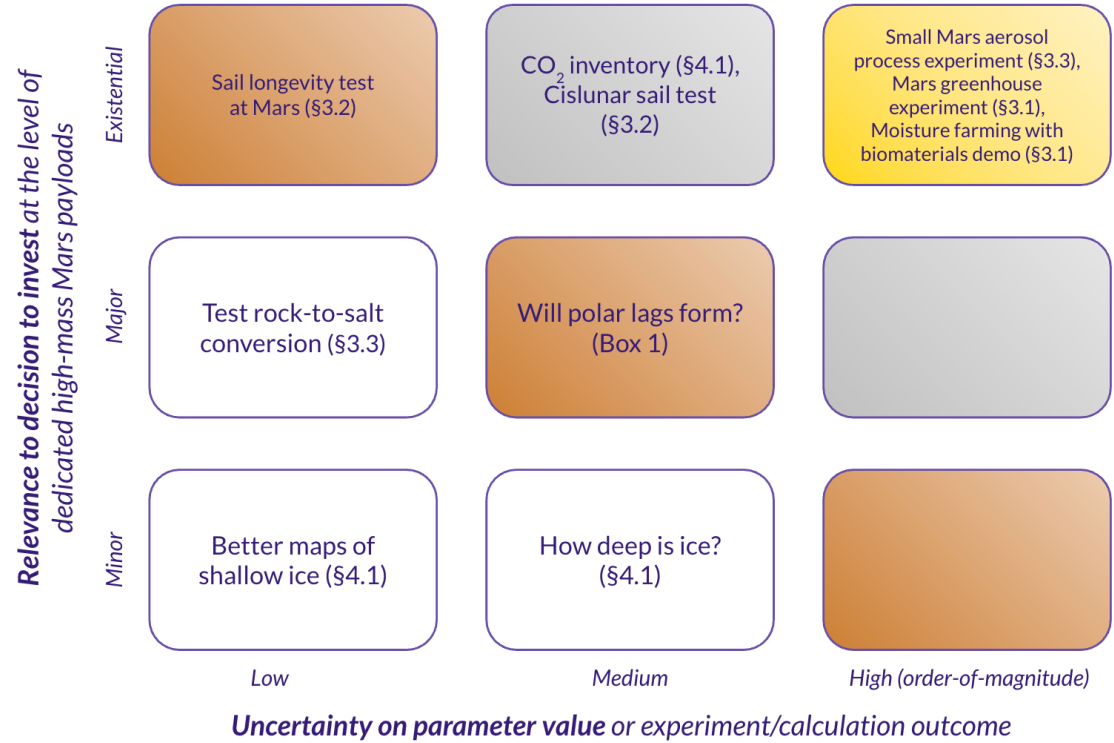

Table 3. Consolidated near-term research options ranked by uncertainty and decision relevance. Large bold letters correspond to relatively inexpensive near-term research questions.

# 4. Synergies

Research into Mars warming would generate knowledge needed for growing human presence in space, regardless of whether warming is ultimately pursued. The converse is also true: accelerating Earth-orbit and lunar activity lowers launch costs and de-risks Mars-relevant technologies [108]. The US (Artemis), China (揽月), and commercial operators plan human landings by 2030, with more than ten Commercial Lunar Payload Services landings planned in 2026-2028. Solar panels made from extraterrestrial soil, if flight-proven [75], potentially cut costs for Mars warming three-fold. Lunar polar ice extraction, if ice quantities are sufficient [109], would validate extraterrestrial water production. Other lunar developments in power systems and metal extraction, if successful, would de-risk their Mars analogs. The basic science required to create a biosphere would also improve long-term management of Earth.

## 4.1. Synergies with basic science needs

Research into the possibility of creating sustainable habitats and biospheres beyond Earth (applied astrobiology) aligns with Mars planetary science [8,10,110]. Mars planetary science and astrobiology gains from resource surveys and process experiments for Mars-warming research. Sites of astrobiological, climate and other geological interest (e.g., polar caps [111]) would need to be investigated before hard-to-reverse changes are made [47]. Sites of historical interest would need to be preserved from destruction (e.g., rusting) in an altered Mars climate. Needs include:

- Better atmospheric science data. A small Mars aerosol release process experiment would answer basic atmospheric boundary layer questions [112], quantify dust-cycle processes (including dry deposition) that are key to forecasting hazardous dust storms [113], support planetary protection by quantifying dispersal of spore-sized particles [114], and verify the dispersibility of Mars-warming aerosols [115]. Electric field data (e.g., Micro-ARES) is needed for aerosol-charging models. The 2025 National Academies consensus report identifies understanding dust storms and aerosol transport as priorities [8].
- Orbiters are needed to monitor natural variability and to relay data to Earth.
- Better maps of subsurface water ice: a potential resource for humans, and microbial habitat on a warmed Mars. The NASA/JAXA/CSA/ASI International Mars Ice Mapper (I-MIM) mission concept addressed this need. Subsurface ice is also the target of the (highly ranked by the Planetary Science Decadal Survey) Mars Life Explorer (MLE) concept.
- According to the 2025 National Academies consensus report, initial human Mars missions would need to include long-duration tests of an integrated ecosystem—"the first time that an integrated ecosystem of humans, plants, animals, and microbes will coexist on a planetary body other than Earth" [8]. This would allow assessing the diversity and productivity of the biosphere that could be supported.

- Soil sample return for life search, and quantification of biocritical trace elements [16] and toxins. China's Tianwen-3 (天问三号) is planned for 2028 [116].
- Measuring adsorbed-$CO_2$ [117] and $CO_2$ ice [118] reservoirs, a key feedback on both natural and deliberate warming.
- Solid-state-greenhouse warming has a near-term water-harvesting application, and as a byproduct also enables a search-for-life experiment. It has been suggested that spores exist at undetectable levels in Mars' soil, waiting for infrequent liquid water. Converting water from ice to liquid using solid-state greenhouse membranes could allow detectable growth, enabling astrobiological research. This is aligned with the Mars Life Explorer mission concept, which was top-ranked by the Planetary Science Decadal Survey.
- Search for deep-buried liquid water [119]: potential habitat for Mars life, or water source for human-base farms.
- Palaeoclimate analysis of existing ice deposits [55,72,111] to quantify climate feedbacks [36].

International coordination, building on groups like the International Mars Exploration Working Group, would make these objectives easier to achieve.

Mars-warming research also synergizes with exoplanet research. Even civilizations that never leave their home planet must eventually deploy climate-altering technologies as their star evolves. Such technologies serve as technosignatures that might be detected across interstellar distances [120].

### 4.2. Synergies with human needs on Mars

The prospect of large numbers of people living on Mars has driven scientific and technologic advance for over 100 years [6,11]. The 2025 National Academies consensus report discusses bioregenerative life support systems for processing waste, refreshing air and water, and supporting food production [8]. Synergies with human needs (Fig. 17) include (in priority order):

1. **Solid-state greenhouse material for warming ice to enable moisture farming solves a near-term in situ resource utilization need.** 0.5 hectares of solid-state greenhouse material could produce enough water for one Starship fuel load (i.e., ~300 tonnes of $CH_4$). This assumes sublimating pore-filling $H_2O$ from soil to 1 m total depth, including 50 cm of dry regolith cover.

2. **MOXIE scale-up automatically generates reduced carbon, which enables Mars-warming and biological applications.** Filling the propellant tanks of (for example) five Starships would take 6,000 tonnes of $O_2$. If propellant production is done by $CO_2$ electrolysis, this generates CO as a byproduct. If this CO were converted to carbon by exothermic CO disproportionation (Fig. 14), this could serve as feedstock for carbon-based IR-resonant materials (e.g., [105]). Scaled-up on-Mars electrolysis testing is already motivated for MW-scale oxygen production for human return missions. A near-term demonstrator of organic carbon production on Mars is feasible (Mars soil has only 0.5 kg/m³ organic carbon [17]); carbon plus nitrogen can contribute to fertilizing soil. Reduced carbon can also be used for bioplastics; natural acetogens can use CO as sole carbon and energy source, while engineered lithoautotrophs use $CO_2$ and $H_2$ (the latter from water electrolysis) for biopolyester production, in both cases without organic carbon supplementation.

3. **Local microbial habitats could potentially produce $O_2$ for human use.** A crater-spanning human habitat (2-km diameter) with 10,000 people [9] needs 4,000 tonnes/year of oxygen (mostly for breathing), which could in principle be provided by a 8 km² photosynthetic habitat. Unlike previous test facilities that are closed-loop, such as Lunar Palace 1 (月宫一号), Europe's MELiSSA, and China Exo-Ecosystem Space Experiment (中国空间站地外生态系统模拟实验) on the Tiangong (天宫) space station [26,121,122], this would involve a large-scale greenhouse exposed to the ground.

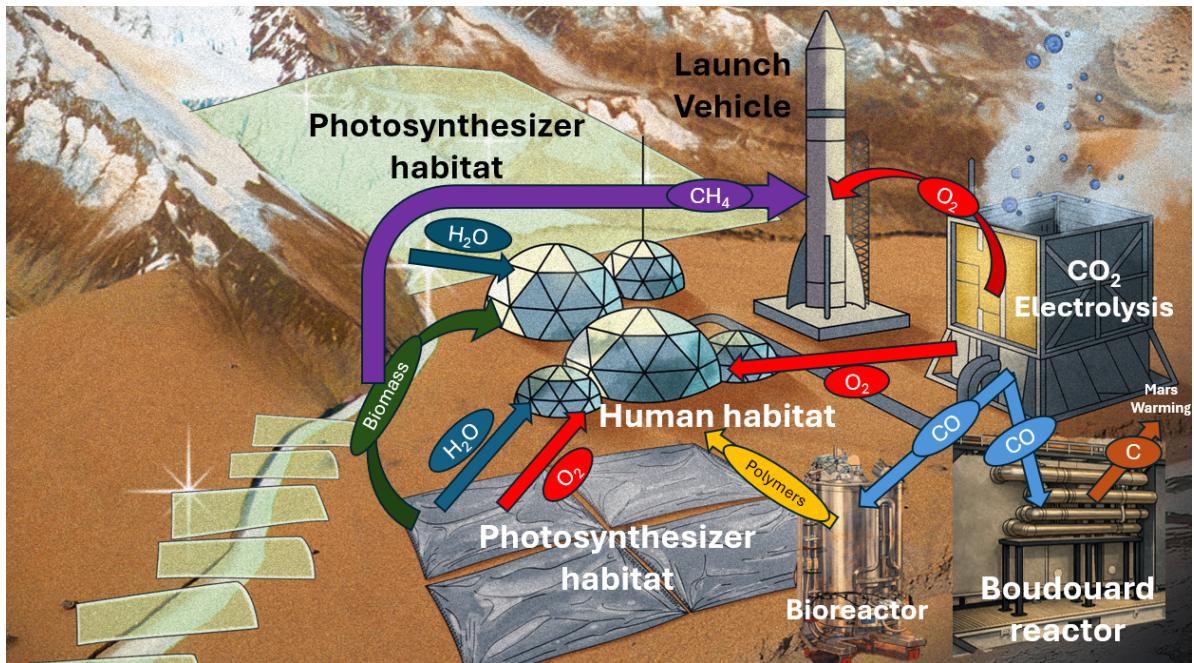

Fig. 17. Example synergies between a human habitat, a microbial habitat, a scaled-up version of the MOXIE experiment, a carbon factory and a launch vehicle.

4. **Microbial communities under membranes could produce oxygen and biomass supporting human bases and atmospheric warming.** Because microbes are hardier than humans, their habitat is relatively cheap—potentially as simple as unrolling an insulator over soil and burying the edges. These reduced requirements include lower pressure, lower temperatures (frozen solid in winter, highs of ~293K) and perchlorate exposure. Some microbes can bioremediate soil under Earth conditions, e.g., metabolizing perchlorate while producing oxygen; whether this extends to Mars soil is untested [18,27,123]. Using Earth arctic tundra productivity [124] as a rough upper bound, if all net primary production were harvested, 1 km² of habitat could make 500 tonnes of biomass; if 20% were edible with reasonable caloric density, this would support 100 people. Inedible biomass could be a backup energy source within oxygenated human habitats, or could be processed to make particles for atmospheric warming. Exponential production of photosynthetic habitat could greatly reduce the costs of membrane production [29]. For biologically produced solid-state greenhouse membranes, if exponential production is possible then it would take >10 years for exponential production to cover >1 km² areas, but this would need relatively little power [29].

> Alternatives: Dark food systems use sunlight to make hydrocarbons that are fed to autotrophs, achieving higher energy efficiency than photosynthesis [125-126, 131]. Extraterrestrial artificial photosynthesis has been demonstrated by Chinese researchers using samples returned from the Chang'e-5 (嫦娥五号) lunar samples [127]. However, unlike biology, dark-food systems cannot self-replicate.

## 4.3. Surface warming and atmospheric thickening together promote conditions favorable for shallow subsurface liquid water

Today, Mars air is too thin to allow pure liquid water in the highlands. Brines with lower melting points may be possible. In the lowlands, evaporative cooling prevents surface liquid water [72,128], and liquid water at the surface would boil at ~280 K. Warming and atmospheric thickening work together to broaden the area of Mars with temperatures and pressure suitable for shallow subsurface liquid water (Fig. 18).

If Mars was warmed, polar $CO_2$ ice would turn into $CO_2$ gas. $CO_2$ loss to space is negligibly slow [129], so atmospheric pressure would rise by at least 2× [117], and possibly >10× depending on the adsorbed-$CO_2$ content of regolith and the $CO_2$ ice content of the polar caps in excess of the radar-confirmed massive $CO_2$ ice deposits [59,117-118,129,130]. However, well-mixed greenhouse agents do not warm Mars' poles efficiently, so this could take $10^3$ yrs [14,20] unless orbiting reflectors accelerate $CO_2$ sublimation (Appendix D).

If warming enables photosynthesis, $O_2$ would build up over many centuries, allowing more complex forms of life [22]. Hybrid or abiotic methods [131,132] could accelerate this. Feedstock is not limiting: the $O_2$ in photosynthesis comes from $H_2O$, which has a Mars column abundance of >35,000 kg/m², much more than the amount of oxygen required to produce a 0.1 bar atmosphere. Water loss rates to space are diffusion-limited and are geologically slow (no significant water escape on civilization-relevant timescales). However, currently unanswered questions include the needed amount of chemically inert "buffer" gas (and from where this could be sourced) in order to prevent uncontrolled combustion, and whether Mars has sufficient accessible electron acceptors (such as $Fe^{3+}$, $SO_4^{2-}$, and $CO_3^{2-}$) to take up the H corresponding to $O_2$ build-up from photosynthetic water-splitting.

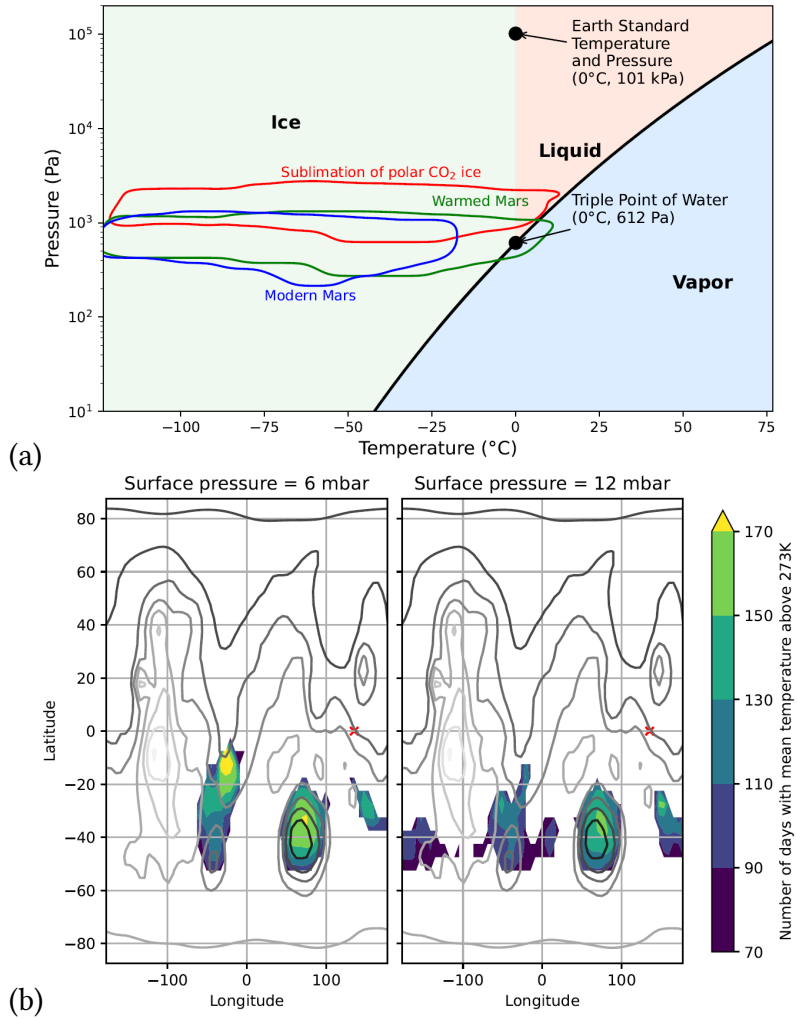

*Fig. 18: (a) Phase diagram for $H_2O$, with the annual distribution of diurnally averaged temperatures in the case of (blue) modern Mars, (green) Mars following 5 years of global warming and (red) Mars following global warming and the thickening of the atmosphere following sublimation of the polar $CO_2$ ice caps. (b) Marginal-melting scenario 5 years following the start of 30 L/s equivalent nanoparticle release from a source at the Equator (corresponding to the red cross), neglecting the temperature effects of radiative cloud feedback. As pressure rises from 6 mbar to 12 mbar due to vaporization of the $CO_2$ in the South Polar Cap, the part of Mars with both surface pressures above the triple point of water and warm-season average temperatures >273K (the blue-green zone), widens.*

# 5. Risks and how research can address them

Table 4. Likelihood-consequence risk framework. Numbered risks are discussed below.

If people reach an agreement to create a large ecosystem on Mars and an international framework to peacefully cooperate to achieve this goal exists, the effort would still involve many risks. Reducing these risks would involve research on fundamental questions that are currently unanswered, discussed below.

Living beyond Earth demands learning from Earth's history [133], including how agricultural intensification, historically, has caused soil degradation, and using extraterrestrial resources sustainably [5]. It is possible that microbial life on Mars subsists deep underground [134], and could be outcompeted by Earth life. This risk might be reduced by never sending humans, as people "will inevitably introduce orders of magnitude more terrestrial microorganisms to Mars than robotic missions have done or will do" (NASA Planetary Protection Independent Review Board [48]). However, in the context of human exploration, the exploration of Mars' surface for existing life will continue [8]. Supposing that the number, quality and breadth of negative results in a search for existing life on Mars all increase, our degree of certainty that Mars is truly as lifeless as it appears to be would increase [135]. Humans have already gone through this process for Earth's Moon.

Top technical risks needing research now ("red brick walls", Table 4) are:

      1. **Would warming actually allow photosynthesis?** Even warm Mars soil is harsh: salty, perchlorate-laden, with high UV fluxes, low atmospheric pressure [15,136] and large diurnal temperature swings [137]. Membranes can shield life from most stressors, making this manageable for local warming. Earth microbes can survive every one of these stressors individually, but no known extremophile can grow in open soil in the combination: advances would be needed to engineer a pioneer micro-organism [138]. Moreover, the climate would need to be optimized for life, not just liquid water. Advances in understanding of biology's potential are needed to design suitable habitable environments. (High consequence, high likelihood).

      2. **Waste stream management.** Mars-warming materials must not impair human or microbial health when tracked into habitats. They should break down into biocompatible,

climate-neutral substances—in the case of engineered aerosol, possible approaches include spacers, thinning to accelerate oxidation, and water solubility. In the short term, laboratory experiments would be needed to find the oxidation time for candidate warming particles under Mars-like atmospheric conditions. Particles would also need to avoid destructive interference with surface hardware. (High consequence, high likelihood).

**3. (For engineered aerosols) Scale-up bottlenecks.** None of the specific particles in [20] are currently mass-produced, and scaling from lab to full production typically takes over a decade on Earth—longer for Mars, given the requirement for space-flight qualification. Reducing this timescale to 5–10 years would assume parallelism (simultaneous scale-up of the Earth engineering model and the flight model), adding to risk. Heavy consumable requirements, if unavoidable, could also break the economics. These long lead times are a key reason why—even though regional-or-larger warming is a longer-term possibility—research should start now, in parallel with quicker-payoff local warming approaches. Work on lunar resource use can be transferred to Mars processes, though this will only accelerate progress if currently scheduled lunar resource use tests go as planned. Some uncertainty will be resolved by particle downselect, since factory requirements are process specific. Work in the next 3 years should include factory architecture studies to better understand risk profiles. (High consequence, high likelihood).

Other risks are lower priority:

**4. Would adding energy to Mars give a warm, sunlit surface?** A warmer Mars might be dustier [67], blocking sunlight—a potential failure mode. We currently do not know how quickly natural aerosols would sweep up engineered aerosols, nor how quickly engineered aerosols would clump at dispersal; if either is fast, aerosol-based warming could be prohibitively costly, favoring gas warming [11,65]. If clumped aerosols cool rather than warm and are easily re-lofted, the net effect could be cooling. Over centuries, rising oxygen could also make the dusty surface and atmosphere explosive. Modeling would be needed to address all these possibilities, in addition to better data for the charge state of natural Mars dust (High consequence, medium likelihood)

**5. Deployment is harder or costlier than expected.** Mars transport prices will stay high if only one beyond-Earth-orbit provider exists. Warming Mars would realistically require government funding, directly or indirectly, but a competitive near-Earth space economy would be needed to bring costs down. Frequent equipment failures, if encountered, would also drive up costs. (High consequence, medium likelihood)

**6. Sustainable resource use.** Resources should be drawn down at rates maintainable for well over $10^4$ years. One risk is slow water loss via downward percolation if heating is excessive. Another key resource is societal coherence to maintain critical systems, as for the industrial fixation of nitrogen on Earth. (Medium consequence, low likelihood)

**7. Reversibility.** Once lives depend on Mars-warming technology, reversibility becomes undesirable. However, the test phase should be reversible by design. This suggests a staged approach might be warranted: hard-to-reverse steps (such as releasing Earth-derived microbes into large warmed regions) should wait until society has committed to a large Mars population. (Medium consequence, medium likelihood)

**8. Can off-Earth manufacturing be tele-operated or autonomous?** Mars factories would need to operate with minimal direct human intervention, requiring autonomous or tele-operated robotics not yet standard on Earth that go well beyond the capability of the "Canada hand" on the International Space Station, the Autonomous Exploration for Gathering Increased Science (AEGIS) system that commands the rock-ablating lasers on NASA's Mars rovers, or the recent use of a Large Language Model to command a 0.4 km drive by a NASA Mars rover. Once astronauts are at Mars tele-operation will be possible with minimal light-travel-time delays. Although lunar soil experience will help, Mars soil would pose novel challenges. (High consequence, low likelihood)

**9. Can humans flourish off Earth?** It is unknown whether human civilization can thrive off-Earth. Radiation is expected to be a hazard at least until atmospheric pressure exceeds 0.1 bar [139]. Given the unknown effects of lower Mars gravity on reproduction, on-Mars tests of animal multi-generational biology are prioritized by the National Academies science strategy for initial human missions [8]. If in the future crew were lost and there were no obvious short-term financial benefits to exploration, society might cease to pay the high costs of sending people to space. Global warming methods would also alter Mars' sky—diffusing sunlight (aerosols) or producing multiple apparent suns (orbiting reflectors)—widening the psychological gap with Earth. (High consequence, medium likelihood)

**10. Geological negative feedbacks.** Summer ice-melt that refreezes in winter creates permafrost cycling (similar to central Alaska), complicating civil engineering. Some outgassed $CO_2$ would be re-adsorbed by soil [117]. A thicker atmosphere eases landings but strengthens winds, stirring up dust and causing greater stress on human-built structures. Engineered aerosols would block some sunlight, reducing solar panel output. Reduced minerals can consume photosynthetic oxygen via oxidation (although this process is slow). Biosphere 2 did not achieve oxygen mass balance—specifically, concrete carbonation indirectly drew down oxygen—showing that unacknowledged solid reservoirs can disrupt mass balance in an ecosystem. (Low consequence, medium likelihood).

# 6. Alternative architectures that could radically lower costs

Significant reorientation of existing space expenditure would be needed to globally warm Mars if it costs $1 bn/K/yr. At this rate, warming Mars would cost ~¼ of 2026 total space expenditure by governments, or 6% of total space expenditure. Lowering costs could bring warming within reach of individual organizations. These ideas are not included in our main calculations but merit exploration during on-Earth research. The top three priorities are:

**Longer particle lifetime.** Engineered-aerosol warming cost would be inversely proportional to particle lifetime. The effective lifetime would be set by how fast aerosols settle onto the surface, how easily they are re-suspended, and how quickly they break down. Photophoretic lofters or particles engineered for ease of re-suspension [107,140-141] could increase effective lifetime, but must overcome strong winter-hemisphere downdrafts. Particles made on one of Mars' moons would quickly spiral down to the top of Mars' atmosphere, potentially lengthening atmospheric lifetime relative to bottom-of-atmosphere release. These hypotheses could be tested with climate modeling and Mars-chamber experiments.

**Biological production of $H_2O$-impermeable, surface-warming habitats.** Growing habitats from biological materials would tap into biology's ability to self-replicate, potentially enabling rapid greening of Mars' surface [142] (Fig. 19). Hybrid approaches using bioplastic appear achievable in the shorter term [143], and would unlock exponential growth [29].

**Elimination of polymer extraction from the habitat replication cycle.** Current bioplastic habitat concepts assume that polymer must be extracted and purified before fabrication (Fig. 19) [29,142-143], adding infrastructure complexity. In highly accumulating strains where polyhydroxyalkanoate (PHA) can exceed 80% (of dry cell mass), biomass itself can exhibit thermoplastic-like bulk properties [144], making direct use of biomass as structural composite material a potentially simpler alternative. Whether mechanical integrity at relevant scales is achievable is an open research question, depending in part on whether (co-)polymer composition can be tuned toward materials with sufficient ductility and interfacial adhesion to cohesively bind adjacent biomass units.

Lower priorities, because they are harder to prove out at low expense, are discussed in Appendix B. One possibility could be investigated using only computer modeling (relatively inexpensive):

**Self-stabilizing feedbacks.** Self-sustaining warming thresholds might be found via experiment, modeling, or palaeoclimate analysis [54-57,111]. More broadly, bootstrapping from biology would need a biologically producible gas with long photochemical lifetime and strong greenhouse potential [63]—from microbes engineered to be non-viable on Earth—and would require large membrane-covered areas for photosynthesis.

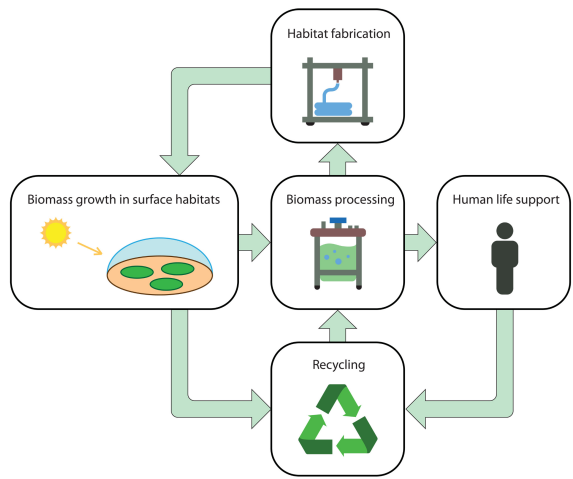

*Fig. 19. (Reproduced from [29]). "Schematic of the biomaterials approach to life support in extraterrestrial environments. In this approach, raw materials for habitat construction are synthesized from in situ resources." Surface habitats made from recyclable bioplastics have a key role.*

# 7. Consolidated roadmap and downselect criteria

## 7.1. Key decision points, sequencing and fallbacks

The complementary strengths of the three warming approaches suggest a potential portfolio strategy; initial research would need to determine what approaches work and at what scales. Our current best guess is shown in Fig. 5.

Three key branching points shape this roadmap (Fig. 20, Tables 5-6). First, particle downselect: Mars-warming particles would need to be long-lived, easy to make, and degrade harmlessly. If no particle achieves this, the aerosol approach fails at global scale, narrowing the roadmap to regional warming, orbiting reflectors, and local membranes. Second, sail design: if no flyable-to-Mars sailcraft can achieve ≤20 g/m², reflectors can only provide small-scale warming—though since paper designs can probably meet this target, rejection would likely require near-Earth flight tests. Third, on-Mars manufacturing viability (which might require on-Mars data): if aerogel or membrane production on Mars cannot achieve break-even relative to Earth shipping within 5 years of production, local warming would stay small. If results are negative, the corresponding pathway would be narrowed or closed.

| Key Decision Points (KDPs) ▲ | When? (central estimate) | Information required/criteria | Minimum funds committed to get to KDP |
|---|---|---|---|
| Particle downselect ▲ | Year 1 | Simulations, lab tests. Most important criterion is >⅓ year particle effective lifetime. | <$1 mn |
| Solar sail system analysis ▲ | Year 2 | Sail design can/cannot fly to Mars and direct sunlight to small surface target for ≤20 g/m² mass using inexpensive materials. | ~$2 mn |
| Selection (or not) of small-scale hosted payloads ▲ | Year 2 | [Local] Projected warming to melting point for >200 m² of soil per tonne of membrane. [Aerosol release] Projected release of 80% of particles without clumping to 1 km distance. | ~$1 mn per payload |
| Go for build of pilot factory ▲ (if regional-to-global track chosen) | Years H+0 to H+2.5 | Detailed design, good kg-scale results. | ~$50 mn |
| Go for climate experiments ▲ (if regional-to-global track chosen) | Years H+5 to H+10 | Results from pilot factory. Output from climate models. | Several billion dollars |

Table 5. Key decision points. H = Decision made for a human base on Mars.

Fig. 20. A research roadmap for assessing the feasibility of warming Mars. Shared milestones, phase boundaries, and decision points marked. H = Decision made for a human base on Mars.

|  | Solid-state greenhouse membranes (§3.1) | Orbiting reflectors (§3.2) | Engineered aerosols (§3.3) | |
| --- | --- | --- | --- | --- |
|  |  |  | Air as feedstock | Soil/rocks as feedstock |
| *Would it warm Mars?* | Yes | Yes | Maybe | Yes |
| *Maturity* | Low | Relatively high | Relatively high upstream, low downstream | Low |
| *Supports a 1,000 person base?* | Yes | Probably | Synergizes with, but does not directly support | Over-engineered for this purpose |
| *Scalability to >$10^6$ km² of warmed area?* | No | Enhances (via pressure increase, Appendix D) | Yes | Yes |
| *Synergy with human needs* | Yes | Plausible but unproven | Yes | Over–produces metals |
| *35 K warming achievable given 0.3-3 bn/K/yr cost tolerance?* | Only for <<$10^6$ km² areas, without exponential production of warming membranes | No, unless spacecraft are made at Mars' moons | Yes for medium cost tolerance [156], but major uncertainties in radiative efficiency and particle lifetime | Yes for medium cost tolerance [156], but major uncertainties in particle lifetime and manufacturing scale-up |
| *Power needed* | << 1 GW | N/A | ~500 MW for ~5 K warming | ~1 GW for ~5 K warming |
| *Mass arriving at Mars* | 3000 tonnes/km² (for aerogel) | $1 \times 10^4$ tonnes @10 g/m² for $10^3$ km² | $2 \times 10^4$ tonnes excluding power system, ~5 K warming | $1 \times 10^4$ tonnes excluding power system, ~5 K warming |
| *Risk associated with uncertainties in physical/chemical properties/processes* | Relatively low | Relatively low to medium | High | High |

Table 6. Summary of Mars-warming approaches. All approaches carry significant uncertainties that research would need to address.

Local warming approaches (such as moisture-farming membranes) could be implemented first as they could offer near-term payoffs for human explorers, in parallel with tests of longer-lead-time approaches.

## 7.2. Cost-to-orbit gating

If results from research on Earth and in space are positive, a decision to go for tests would depend on the trajectory of cost to orbit. If launch costs fall fast toward $100/kg, a pilot factory would be a reasonable path forward. If launch costs remain ~$1,000/kg, longer-lead research into Mars-system production (Fig. 21, §6) would likely be preferable.

The results from the research phase could all be negative, and if extant life were discovered on Mars (which would be a very important scientific finding) then terraforming would be put on long-term pause. In either case, near-term research would still advance the science of sustainable habitats and biospheres, which is relevant to Earth climate modeling, asteroid resource utilization, and any future off-Earth settlement on a solid body.

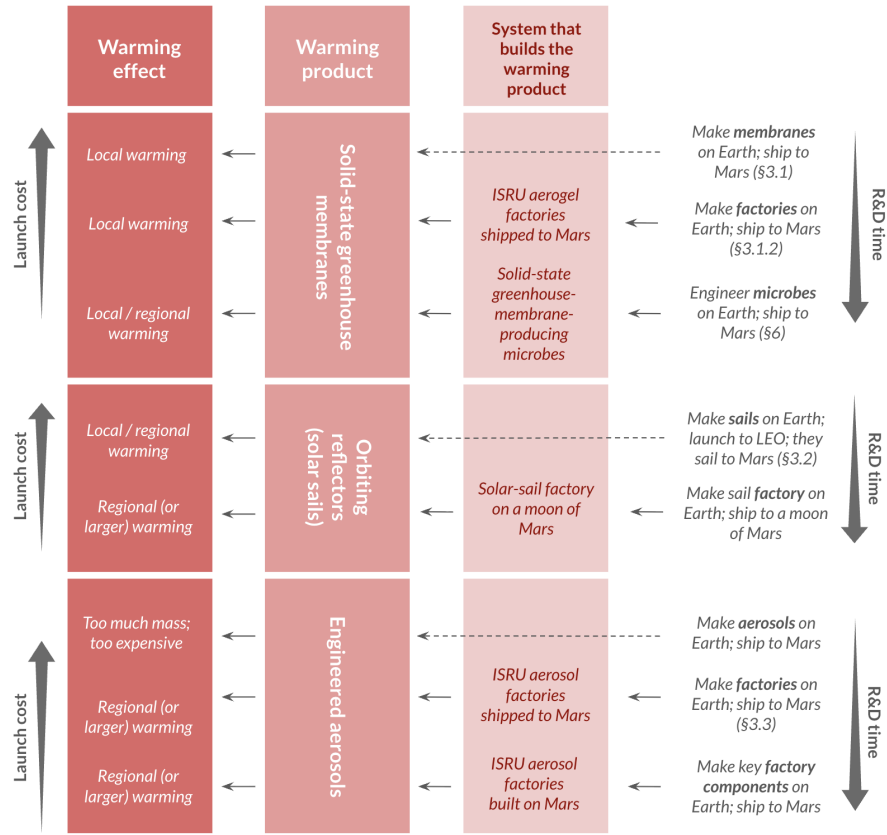

Fig. 21. An overview of the tradeoffs between longer R&D time and lower launch costs for different ways to implement each Mars-warming pathway (§7.2). If launch costs fall fast toward $100/kg, it makes sense to produce warming agents on Earth and ship them to Mars (large mass-to-Mars → higher launch costs). If launch costs per kg remain high, it makes sense to invest in R&D to develop on-Mars warming-agent-manufacturing capability (lower mass-to-Mars because warming agents will be produced with Mars-sourced materials → lower launch costs).

### 7.3. Cross-cutting infrastructure

Cross-cutting infrastructure needs align with existing basic science and astrobiology needs: Mars telecommunications and climate monitoring, Mars environmental test chambers, climate models with aerosol–climate feedbacks, AI, and autonomous robotics (e.g., NASA's STRIDE program), and power systems.

The applied astrobiology research agenda [10] spans disciplines that do not currently have a shared funding home. Historical precedents suggest that building such a field—as with astrobiology in the early 2000s, through organizations like the NASA Astrobiology Institute— requires dedicated programmatic support. This research would not emerge organically from short-term mission life support—astronaut safety rules out staking lives on experimental ecosystems. Integrated-ecosystem tests at Mars' surface are given a higher priority by the 2025 National Academies consensus study [8].

In summary, as launch costs fall, human choices might determine Mars's environment just as they determine Earth's (including the choice to set aside parts of Earth as wilderness) [3,44]. If the cost of access to space falls towards ~$100/kg, Mars-warming research would be the critical path for a second world-scale biosphere.

## 8. Conclusion

This roadmap identifies three approaches to warming Mars and their research needs. Solid-state greenhouse membranes offer the nearest-term benefits, with direct applications to moisture farming and life support at human bases. However, they do not scale to large fractions of Mars' surface area without on-Mars manufacturing. Orbiting reflectors could warm intermediate-scale areas ($10^2$–$10^5$ km²) if sailcraft areal mass can reach ≤10–20 g/m², roughly 3× below the current state of the art. Engineered aerosols might warm at regional-to-global scale, but face unresolved questions about particle lifetime, production scale-up, and biocompatibility. No approach has been shown to be simultaneously affordable, safe, scalable, and to enable extending life beyond Earth; the research would determine which, if any, can do so.

The gating decisions that would shape the next phase—particle downselect, sail system analysis, and go/no-go for small-scale hosted payloads—can be reached for a combined investment of ~$5 mn within three years (Table 5). Negative results at any gate would narrow the roadmap and redirect effort toward the remaining approaches. A finding that no approach is viable would itself be valuable, as it would constrain planning for long-term human presence on Mars, and would inform the broader question of where beyond Earth, if anywhere, it would be possible for large numbers of people to self-sustain.

The proposed research advances Mars atmospheric science, aerosol microphysics, in situ resource utilization, ecosystem science, and climate modeling regardless of whether warming proves feasible—all priorities identified by the 2025 National Academies science strategy for human Mars exploration [8]. Similarly, hosted payloads are process experiments that generate prioritized science regardless of the outcome of the corresponding technology demonstration. If early results are positive, they would provide the quantitative basis for government-scale programs to evaluate whether extending habitable conditions beyond Earth is achievable, at what cost, and on what timescale.

Even under optimistic assumptions, warming at kilometer scale is at least a decade away, and wider environmental modification would require sustained investment over many decades beyond that. These long timescales are a reason to start foundational research now—but not to mistake early-stage feasibility studies for near-term capability.

## Acknowledgements

We thank Y. Takubo, T. Nakagawa, P. Buhler, L. Coffin, J. Haqq-Misra, T. Vora, and M. Nielsen for advice, ideas, calculations, and for reading drafts. We thank all participants in the Green Mars workshops, without implying agreement with all points made here. This work was funded in part by Astera Institute.

# Appendix A

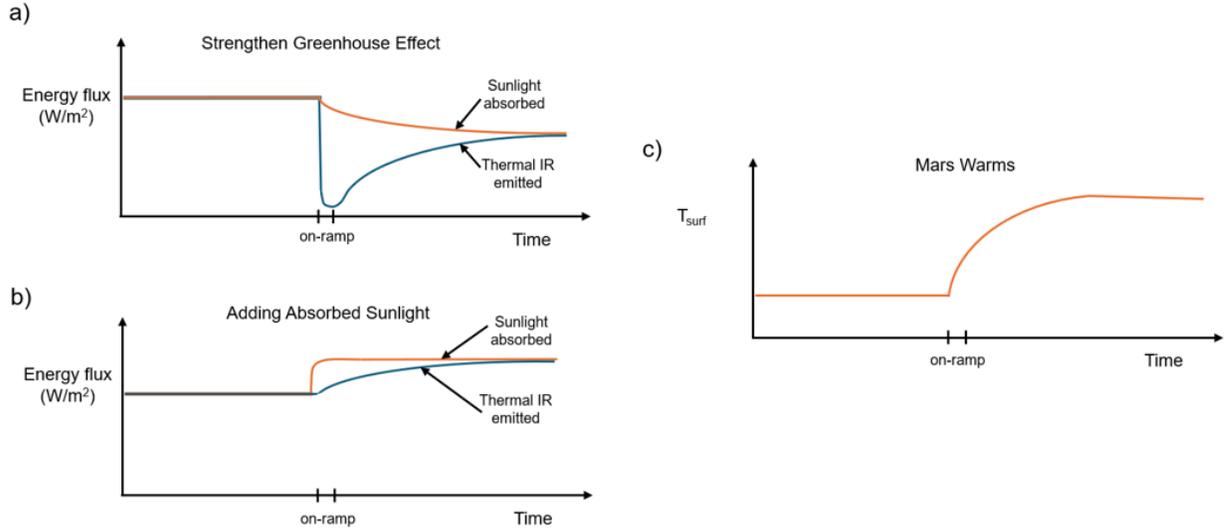

Fig. A.1. How Mars responds to changes in planet energy balance. For panel (a), we assume this greenhouse agent makes the planet more reflective in the visible.

### Timescale for Mars' surface temperature to change (§1.2)

If Mars' surface were warmed overnight, then assuming a large, 10% energy imbalance increased infrared emission would cool it back to 210 K within ~5 days. Mars' surface radiates to space as $F_{emitted} = \varepsilon \sigma T_{surf}^4$, where $\sigma$ is the Stefan-Boltzmann constant, and emissivity $\varepsilon \approx 1$, because Mars' natural emissivity-reducing greenhouse effect is weak (Fig. 2).

$$\tau \approx \frac{\Delta T}{F} \left( \underbrace{c_{p,CO_2} \frac{P_{atm}}{g}}_{C_{atm}} + \underbrace{\rho_{soil} c_{p,soil} \sqrt{\frac{\kappa_{soil} \tau}{\pi}}}_{C_{soil,eff}(\tau)} \right) \approx 5 \text{ days} \quad (1)$$

This timescale is set by the heat capacity of the thin atmosphere ($C_{atm}$) plus the part of the soil that responds to the forcing ($C_{soil,eff}$), where $\Delta T$ = 20 K, $F$ = 11 W/m² (10% of planet-averaged absorbed sunlight), $c_{p,CO_2}$ = 750 J/kg/K, $P_{atm}$ = 610 Pa, $g$ = 3.7 m/s², $c_{p,soil}$ = 1000 J/kg/K, $\rho_{soil}$ = 1500 kg/m³, $\kappa_{soil}$ = 4 × 10⁻⁸ m²/s (from Insight heat-flow-probe data), and we use a single-column average as representative of the planet.

### Why the minimum mass of aerosol needed to warm Mars is approximately 3 million tonnes (§1.2)

Consider an atmospheric layer of gas or aerosol. The layer is transparent to sunlight but has optical depth $\tau$ = 1 (dimensionless) at wavelengths 5-50 μm, where Mars radiates to space. The amount of infrared radiation that escapes to space is

$$I = I_0 \exp(-\tau),$$

where $I_0$ is the upwelling infrared radiation at the surface. For infrared light moving vertically upward, $\tau = \kappa_\lambda \, m_c$ (kg/m²) where $\kappa_\lambda$ (m²/kg) is the wavelength ($\lambda$)-dependent opacity, and $m_c$ (kg/m²) is column mass. The optical depth (for spherical particles with radius $r$) is

$$\tau = 3 \, Q_\lambda \, m_c / 4 \, r \, \rho,$$

where $\rho$ is particle density and $Q_\lambda$ is the absorption cross section at a given wavelength divided by the geometric cross section. $Q_\lambda$ averaged over Mars' emission spectrum can be >10 for small, conductive particles that resonate with light of the given wavelength [24]. To warm Mars, we want high $\kappa$ in the Mars-emitted wavelengths, and low $\kappa$ at solar wavelengths. $\tau = 1$ is the threshold for major alterations in planet energy balance (corresponding to tens of K change in Mars' surface temperature). For $\tau = 1$, the mass in the atmosphere is roughly

$$m_{particles} > 2 \, r \, \rho \, A \, / \, Q = 3 \text{ million tonnes.}$$

where $\rho$ = 2,000 kg/m³, $r$ is particle radius (>50 nm), $A$ is planet area (1.4 × 10¹⁴ m² for Mars), and $Q$ = 10.

### Energy needed to add volatile-rich objects to Mars is unreasonably high (§1.2)

To warm Mars by adding volatile-rich objects to the atmosphere, objects from ≥2.8 AU would be needed. Even setting aside the safety concerns regarding moving asteroids, the power budget needed is too high. To add even 10% to Mars' atmospheric mass of $M_{atmos}$ = 2.5 × 10¹⁶ kg by directly altering the orbit of an asteroid requires a sustained power of approximately

$$P = \frac{M \, \varepsilon_{\Delta v}}{\eta \, t} = \underbrace{(0.10 \, M_{\text{atmos}} / f_{\text{useful}})}_{\text{required delivered mass}} \varepsilon_{\Delta v} / (\eta \, t) \approx 1 \times 10^{13} \text{ W}$$

where $\varepsilon_{\Delta v}$ is the energy per unit mass for the velocity change of ~0.5 km/sec from a position in the Trojan swarm to a decade-timescale journey to a Jupiter gravitational slingshot, $M_{atmos}$ is the mass of Mars' atmosphere, $f_{useful}$ is the fraction of an asteroid's mass with warming-useful volatiles (assumed optimistically to be 0.1), $\eta$ is the fraction of the propulsion system's energy that translates into $\Delta v$ (assumed to be 0.3), and $t$ is 35 years (chosen to match the build-and-test period assumed in §3). This ignores photochemical losses of $NH_3$, which would require greater imports. Although solar reflectors volatilizing comet ice to act as a natural thruster might in principle reduce the power requirement, a more practical way of warming Mars is to use the atoms already present on Mars.

### Controls on particle lifetime (§3.3)

Particles can collide and stick via Brownian diffusion or differential settling. The ultimate loss channel is dry deposition (impaction, interception, Brownian diffusion, gravitational settling), offset by re-lofting. The timescale, $\tau_c$, for the particle concentration, $N_0$, to halve from its original value for a constant coagulation kernel $K$ (m³ s⁻¹) is [145]

$$\tau_c = \frac{2}{K \, N_0}$$

$K$ is $O(10^{-15})$ m³ s⁻¹ for spheres on both Earth and Mars, but increases twenty-fold for high-aspect ratio particles. The generalized form of the coagulation coefficient due to Brownian motion when particles are not charged is [145]

$$K_{12} = 2\pi(D_{p1} + D_{p2})(D_1 + D_2)$$

where $D_1$ and $D_2$ are particle diffusivities, $D_{p1}$ and $D_{p2}$ are the particle diameters, and we have neglected a coefficient that is close to 1 for the aerosol diameters considered. When particles are charged, the coagulation kernel changes. Clumping also occurs via differential gravitational settling and the associated coagulation coefficient is

$$K_{12}^{GS} = \frac{\pi}{4}(D_{p1} + D_{p2})^2(v_{t1} - v_{t2})E(D_{p1}, D_{p2}),$$

where $D_{p1}$ and $D_{p2}$ are the particle diameters, $v_{t1}$ and $v_{t2}$ are settling rates (m/s) and $E$ (a function of $D_{p1}$ and $D_{p2}$) is the collision efficiency. As a reference, Mars dust settling rate is mm/s, while all particles considered here have settling rate below 1 mm/s (some orders of magnitude less). The overall (summed) collision kernel is then used to determine $\tau_c$.

Depending on the dust loading and collision efficiency, $\tau_c$ is on the order of hours to days for graphene without mitigation, and weeks for metal aerosol. If the dust has more than 30 charges, as predicted by the most sophisticated currently available model [103], the clumping timescale for metal aerosol is months to years. The coagulation kernel is divided by a ratio, $W$, to take account of particle charge effects. The key ratio is $W = (e^\kappa - 1) / \kappa$, where $\kappa$ is a ratio of energies:

$$\kappa = \frac{z_1 z_2 e^2}{4\pi\varepsilon_0 \varepsilon (R_{p1} + R_{p2})kT}$$

where $z_1$ and $z_2$ are particle charges in units of elementary charge $e$; $r_1$ and $r_2$ are particle radii; $\epsilon$ is the relative permittivity ($\approx 1.0005$); $\epsilon_0$ is the permittivity of free space; and $k$ is the Boltzmann constant. Graphene disks have a shorter clumping timescale than aluminum ribbons because more individual particles are needed for a given column density.

*Details of the engineered-aerosol warming approach (§3.3)*

Using Mars air as feedstock:

To produce sufficient graphene flux on Mars to warm Mars by 35 K over 35 years with N-doped graphene nanodisks, approximately 200,000 mass-optimized Big Atmospheric MOXIE units [146] would be required. These would intake Mars atmosphere and produce carbon monoxide and oxygen ($2CO_2 \rightarrow 2CO + O_2$). The CO from each unit would be directed into a corresponding Boudouard reactor, which produces C ($2CO \rightarrow CO_2 + C$). The C would be used to make N-doped graphene ($N_2$ would be harvested atmospherically). We assume the Boudouard+doping systems would have a combined mass of 500 kg/unit, and the Big Atmospheric MOXIE systems 400 kg/unit. All together, these would comprise 180,000 tonnes and require ~5 GW of power. At $2,000/kg launch cost to Mars, and assuming

$1,000/kg average hardware procurement cost (excluding solar panels), this warming path would cost approximately $0.6bn / K / year. However, this would only happen if the cost of currently available ultralight (as low as 73 g/m$^2$) solar panels falls to the per-kilogram cost of Earth industrial solar panels ($20/kg). Thus, Mars warming is only affordable in our medium cost tolerance scenario if the dramatic learning curve (cost decreasing with increasing cumulative production) for on-Earth solar power also applies for the imminent large increases in solar panel production for off-Earth use. See [156] for more detail, and see Fig. 13a for a CONOPS sketch of this approach.

There are many alternative routes to using Mars air as feedstock for Mars-warming particles. Graphene could be made by shear exfoliation [147], sonication, or flash Joule heating [148]. Doping could be done with $NH_3$ or by irradiation. Non-graphene organics (including $C_{60}$) have resonances near the ideal Mars-warming wavelengths. Carbon nanotubes can resonate at Mars-warming wavelengths [149], but it is not clear how to suppress their visible absorption.

We assume chemical doping using nitrogen. Mars' atmosphere has 1 nitrogen atom for every 16 carbon atoms, enough for strong doping. N could be added to graphene from $N_2$ by (for example) Plasma Enhanced Chemical Vapor Deposition.

To estimate system cost, we use a scaled-up "mass-optimized" design from [146]: a 400-kg Mars atmosphere electrolysis system that makes ~27 tonnes/Earth-year of $O_2$. We exclude the oxygen liquefaction system (which is not needed for warming Mars; excess oxygen could be vented to habitats). Assuming 500 kg more mass would be needed for each system to account for (a) the conversion of CO to graphene disks and (b) doping the graphene with N, we get 900 kg. In other words, a 0.9-tonne system would make 10 tonnes of C per year. To maintain traceability to existing designs [146], we assume large numbers of this design would be used, although a bigger version would probably be more efficient. The corresponding mass for particle production (atmospheric electrolysis + graphene particle production) needed to double the strength of Mars' greenhouse effect (~4 K global warming), using the assumptions of [20], would be 500 tonnes/year supplied from Earth over a 35-year period. See [156] for more detail including strong warming scenarios. At a $2,000/kg launch cost to Mars and $1,000/kg average non-power hardware procurement cost, and assuming ultralight solar panels (73 g/m$^2$) become available for standard-solar-panel costs of $20/kg, a system capable of producing 2 L/s graphene would cost $2 bn/year (assuming amortization over 35 years, and allowing $20 bn for R&D costs). However, the mapping between 2 L/s and doubling the strength of Mars's greenhouse effect is currently very uncertain due to the open research issues listed in Table 2 and adjacent text.

<u>Using Mars' soil and rocks as feedstock:</u>

To produce sufficient magnesium aerosol flux (3.5 kg/s) on Mars to double the strength of Mars' greenhouse effect (i.e., +4 K) over 35 years (assuming that carbothermally produced

Mg aggregates perform similarly to engineered Mg nanorods on a warming-per-column-density metric basis, or that Mg nanorods can be produced via carbothermal reaction), substantial mining and beneficiation hardware would need to be shipped to Mars. This would be used to process salty deposits rich with $MgSO_4$. The magnesium sulfate would be extracted from the rock using agitated water leach, then heated to drive off $SO_3$. The remaining MgO would be heated further and combined with MOXIE+Boudouard-derived C in carbothermal reactors to get Mg + CO, then rapidly quenched to avoid recombination (e.g., MagSonic process). We estimate the total hardware mass (including solar panels) at ~1 × $10^4$ tonnes [156]. At \$2,000/kg launch cost to Mars, assuming \$1,000/kg average hardware procurement cost (excluding solar panels), and assuming the cost of currently available ultralight solar panels (73-145 g/m²) falls to the per-kilogram cost of Earth industrial solar panels (\$20/kg), a central estimate for the cost of this warming path is approximately \$0.08 bn/K/year, but with order-of-magnitude uncertainties of which the most important is particle lifetime. (See [156] for more detail including strong warming scenarios, and see Fig. 13b for a CONOPS sketch of this approach.)

Mg extraction via carbothermal reduction (MgO + C → Mg + CO) could be done using Mars-derived carbon [150]. Heat would be needed to drive the carbothermal reaction. The carbon could come from electrolyzing $CO_2$. Assuming 50% thermodynamic efficiency, the system would need >30 km² of solar panels. Some efficiency might be gotten by using surface reflectors to direct sunlight to a concentrated solar power carbothermal reduction system. A disadvantage of a metals approach would be the lack of synergy with early human needs, since landed spacecraft might be recycled for metals initially.

Particles might be produced via aerosol-phase approaches. Earth produces >$10^7$ tonnes/year this way, including nano-$TiO_2$ and carbon black [151]. Gas-phase formation commonly yields fractal aggregates composed of sintered primary particles, rather than compact spheres. Such morphologies introduce an additional tunable parameter: the fractal dimension, which describes the degree of aggregate compaction. This parameter strongly influences the optical properties and aerodynamic properties of the aerosols. Elongated, low-density ("lacy") aggregates could in principle be engineered to combine extended atmospheric lifetimes, with respect to 1D materials, and with geometries that approximate dipole-like radiative absorbers. However, particle sizes (~few μm) may be health-hazardous, requiring engineering for rapid degradation in habitats (Table 2).

Metals could alternatively be obtained via Molten Regolith Electrolysis (MRE). MRE is relatively technically mature, but energy-expensive and may require frequent component swap-outs. Another option could be sulfuric acid Fe extraction [152] (low energy but low pH may require frequent component swap-outs).

Size and shape control for Mars-warming aerosol does not have to be precise. As the full-width-at-half-maximum (FWHM) in wavelength is a factor of ~2.5 for blackbody radiation, and resonant wavelength is roughly proportional to size, a size distribution FWHM of a factor of 2.5 (or more) is acceptable.

# Appendix B

*Additional alternative architectures that might radically lower cost (§6)*

This Appendix is a continuation of §6. Many additional ideas are compiled in [153].

*More aggressive Mars-system resource utilization* (Fig. 21). 3D-printing massive factory components on Mars (e.g., Relativity Space) would cut Earth-shipping costs. Making sails at Mars' airless city-sized moons (escape velocity ~10 m/s) by physical vapor deposition could reduce launch costs. One of Mars' moons will be compositionally analyzed by Japan's MMX mission, which launches in 2026. However, knowing moon composition is only one of many steps that would be required for milli-gravity manufacturing, which is unproven.

*Biogenic particles as aerosol feedstock.* Desiccated, non-viable PHA-rich biomass is an unexplored alternative feedstock for engineered aerosols, potentially leveraging biological growth kinetics rather than purely abiotic manufacturing—provided that the infrared absorption characteristics of the polymer matrix and the aerodynamic behavior of the resulting particles (cells 1-2 microns in length when desiccated, or spores ~1 microns in size) prove competitive with inorganic candidates.

*Spacecraft reuse.* The $2,000/kg Mars-transport cost estimate assumes that spacecraft that are sent to Mars never return to Earth. Earth-Mars cycling could reduce costs.

*Use Mars-abundant minerals.* Mars salts absorb at warming-relevant wavelengths and are cheaply extractable from soil, but the required ~30 nm particle size creates severe coagulation challenges that may prove the method to be unviable. Future optical measurements of warming-relevant salts (e.g., $MgSO_4 \cdot nH_2O$, 0.4–50 μm [154]) could revisit this.

# Appendix C

*Willingness to pay.*
Willingness to pay is unknown and will depend on perceived benefits, space economy growth, and inter-governmental coordination. Modifying a local habitat for warming is relatively inexpensive and corresponds to the food requirements of even a small Mars base. To assess whether broader-scale warming research merits further study, it is useful to establish rough cost benchmarks. For Mars surface warming by 35 K, we consider cost tolerance scenarios of $0.3 bn/K/yr, $1 bn/K/yr, and $3 bn/K/yr—corresponding roughly to the amortized cost of a single international civil space project (the International Space Station), one-quarter of current space spend by governments, and one-quarter of total space spend, respectively. Engineered aerosols satisfy this cost tolerance (Tables 6-7) but with major uncertainties in

radiative efficiency, particle lifetime, and/or manufacturing scale-up, depending on feedstock used.

If the human population on Mars grows (many scenarios are possible; see [155] for one proposed scenario), demand for Mars-produced food, oxygen, heat, power, and other resources would increase, potentially incentivizing operators to develop warming systems. The 26-month launch-window cadence would set the pace of iterative improvement. This roadmap models build-up over 35 years (after a decision to permanently inhabit Mars is made), because growth in human numbers and associated demand is unlikely to be faster than this.

This would require a large investment. Our focus is on the smaller, near-term research investments that could determine whether those large investments are justified.

| Parameters varied one at a time: | Currently demonstrated / conservative parameters | Baseline | Floor |
|---|---|---|---|
| Transport costs | 18.66 bn/K/yr ($100,000/kg) | 0.55 bn/K/yr ($2,000/kg)[4] | 0.28 bn/K/yr ($500/kg) |
| Particle effective lifetime | 5.54 bn/K/yr (0.1 years) | 0.55 bn/K/yr (1 year) | 0.11 bn/K/yr (5 years) |
| Producibility of solar panels on Mars | 1.12 bn/K/yr (no) | 0.55 bn/K/yr (yes) | 0.55 bn/K/yr (yes) |

Table 7. Total cost sensitivity to parameters, for +35 K warming, considering air-derived particles. Estimates are current best guess. See [156] for details.

# Appendix D

*Although whole-sail areal density <4 g/m² would likely be needed to keep costs reasonable, illuminating the South Polar CO₂ ice deposit with orbiting reflectors could raise Mars' atmospheric pressure (§3.2/§4.3)*

Enough $CO_2$ ice is buried in the South Polar cap to double atmospheric pressure, but it sublimates slowly unless strongly warmed [130]. Aerosol warming appears to be inefficient at the poles [20,24]; orbiting reflectors might provide strong polar warming. Doubling atmospheric pressure would yield only 5 K surface warming. The effect is semi-permanent (natural $CO_2$ deposition is slow [117]), and positive for surface liquid water availability (Fig. 18).

*How it works:* The residual South Polar $CO_2$ cap (~$10^5$ km², containing $2.8 \times 10^{16}$ kg $CO_2$ ice [130]) might be sublimated using orbiting reflectors. The power $P$ required is:

$$P = E/t = (L_{\mathrm{CO_2}}\, m_{\mathrm{CO_2}})/t \approx 4 \times 10^{13} \text{ W}$$

where $E$ = absorbed energy, $t$ is time (35 years), latent heat of sublimation of $CO_2$ ($L_{\mathrm{CO2}}$) = 590 kJ/kg, and $CO_2$ ice mass $m_{\mathrm{CO2}}$ = $2.8 \times 10^{16}$ kg [130]. The radiative cooling (thermal emission) of $\epsilon_{CO2} \sigma T_{frost}^4$ = 22 W/m² (where $\epsilon_{CO2}$ = 0.8 is the emissivity of the polar $CO_2$ ice, $\sigma$ is the Stefan-Boltzmann constant, and $T_{frost}$ ≈ 150 K) is mostly supplied directly by natural sunlight. The minimum mass in solar sails, $m$, is

$$m_{\text{sail}} = \sigma_{\text{sail}} A_{\text{sail}} = \sigma_{\text{sail}} \frac{P}{I_{\text{Mars}}(1-\alpha)\eta}$$

where $I_{Mars}$ = 590 W/m² is solar energy reaching Mars, $\eta$ is the efficiency of redirection of that energy to a zone of interest ($\eta$ < 1, as the orbiting reflector cannot see the target for the entire orbit, the orbiting reflector has albedo <1, etc.), and $\alpha$ ≈ 0.7 is the albedo of the $CO_2$-bearing zone of interest. As the diameter of the $CO_2$-bearing zone of interest ($D_{ZOI}$) is ~200 km, a reflector with perfect pointing is effective from a distance ($t_{sep}$) from the zone of interest $t_{sep}$ < $D_{ZOI} a_{Mars}/D_{Sun}$ = ~3 × 10⁴ km, where $D_{Sun}$ is the Sun's diameter and $a_{Mars}$ is the distance of Mars from the Sun. For <1000 km-altitude orbits, $\eta$ = 10-20%. Imperfections in the reflector can further lower efficiency. For 1 g/m² and $\eta$ = 5%, this gives $m$ = 2 × 10⁹ kg (2 × 10⁶ km²). Darkening the cap (reducing $\alpha$ to 0.1, for example, with a thin layer of basaltic sand [23]) would absorb 3× more sunlight and lower mass to $m$ = 6 ×10⁸ kg. Airmass losses due to dust are lower over the pole than elsewhere (given that dust is less common at extreme latitudes) and would likely reduce energy delivered to the pole by < 20% [Mars Climate Database: https://www-mars.lmd.jussieu.fr/, taking the ratio between incident solar flux at top of atmosphere to incident solar flux at surface.]

A 1 km² sail at 700 km altitude can reflect light to the region of interest during ~20% of its orbit, yielding a Mars-year-averaged power deposition rate of ≈75 MW (assuming sufficient attitude control to point precisely and station-keep). In order to sublimate enough $CO_2$ ice to give 6 mbar of gas over a span of three-and-a-half decades, a large number of such sails (roughly half a million) would be needed. While this could pose a space sustainability risk, the freedom in the choice of orbital elements at Mars, including altitude, to meet the sun synchronous orbit requirement may help to mitigate the problem. In addition, any space traffic management strategy at Mars will benefit from lessons learned from large constellations in Low Earth Orbit, making the sustainable implementation of a solar sail constellation more likely. Finally, whole-spacecraft areal density would need to be ≤4 g/m² to keep costs below $10bn/year[11].

To warm Mars by 35 K requires ~2 orders of magnitude more reflector area than sublimating the $CO_2$ ice at the South Pole—challenging, and plausibly requiring Mars-moon sail production (Appendix B). A possible location for reflectors for hemispheric warming could be the (solar-radiation-pressure adjusted) Mars-Sun $L_2$ point [87].

---

[11] Assuming $100/kg launch costs to Low Earth Orbit, launch over a 35-year period, 35-year spacecraft lifetime, and $100/kg sailcraft procurement costs. Details at [156].

## Appendix E

*Details of Figure 5.* Figure 5 compares three Mars-warming pathways and their estimated costs as a function of Mars surface area warmed. These are shown using bands to represent uncertainty in cost estimates. See [156] for details.

Solid-state greenhouse membrane (specifically, aerogel) cost estimates are shown in blue. The central estimate corresponds to payload-to-Mars costs of $2000/kg and aerogel-procurement cost of $25/kg. The lower bound corresponds to payload-to-Mars costs of $500/kg. Greenhouse membranes with less mass per unit area could further lower costs. The upper bound is one order of magnitude above the central estimate. A horizontal dashed line at $10B corresponds to the as-yet-undemonstrated case where microbes are engineered to allow (in combination with bioplastic-to-habitat converters, Fig. 19) exponential production of bioplastic habitats that have solid-state greenhouse properties [28,29,142]. In this scenario, a relatively small factory on Mars can generate large amounts of surface coverage.

Engineered aerosol cost estimates are shown in pink. The central estimate corresponds to aerosols made using Mars' atmosphere as feedstock (specifically, N-doped graphene) in a scenario in which payload-to-Mars costs are $2000/kg and non-power hardware-procurement costs are $1000/kg. (Solar panels are assumed to be the power source). The lower bound of the primary pink band corresponds to the case in which payload-to-Mars costs drop to $500/kg and solar panel procurement and shipping costs are excluded (i.e., assuming in situ production of solar panels at small cost). The light-pink strip below the main pink band represents another lower bound to account for the possibility of further mass optimization beyond Big Atmospheric MOXIE [146]. The upper bound is one order of magnitude above the central estimate.

Solar sail cost estimates are shown in yellow. These cost estimates are anchored to three points in three domains. Consistent across each domain, the objective of the sail-induced warming is to "double sunlight" to the region of interest (year-averaged power equivalent to 140W/m$^2$). This is slightly more challenging than for 35 K warming. The three domains are:

- 0 km$^2$ → ~3 km$^2$ (i.e., zero →  a ~1-km-radius human habitat)
- ~3 km$^2$ → ~3 × 10$^4$ km$^2$ (i.e., a human habitat →  the area of the ~100-km-radius CO$_2$ ice deposit at Mars' South Pole)
- ~3 × 10$^4$ km$^2$ → "all" of Mars (i.e., a 100-km-radius region → one hemisphere of Mars)

The minimum spot size of reflected light from a sail orbiting at ~700 km altitude is approximately 2 km. This means that it is no cheaper to double the sunlight reaching a 100 m$^2$ region than it is to double the sunlight reaching a 3 km$^2$ region. The central estimate in this domain assumes 10 g/m$^2$ whole-spacecraft areal density; $100/kg launch costs to Low Earth Orbit, $100/kg spacecraft procurement costs, and $0.5B in R&D costs. The lower bound in this domain assumes launch costs drop to $30/kg. Launch costs for sails are lower than for other methods because sails need only be launched to Earth orbit, from which they can fly themselves to Mars.

The second domain interpolates between the habitat-warming objective and a new objective, sublimating the $CO_2$ ice at Mars' South Pole. Calculations (details at [156]) indicate that ~6 × $10^5$ km² of reflector surface in a sun-synchronous polar orbit would be needed to sublimate the $CO_2$ ice over a period of 35 years. The average power required for sublimation is ~9× higher than that needed to provide 140 W/m² to the same region, so ~6 × $10^4$ km² would be sufficient to achieve the "double sunlight" metric. Doubling sunlight at Mars' south pole would not be enough to double Mars' atmospheric pressure in 35 years, but using this metric allows for apples-to-apples comparison of costs as a function of spatial scale. We anchor the central estimate at the point (~3 × $10^4$ km² , $55 bn) and interpolate between the first domain's central estimate of (~3 km² , $3.3 bn). $55 bn assumes $100/kg launch costs, $100/kg procurement costs, and 4 g/m² spacecraft areal density (reduced from the human-habitat assumption of 10g/m²). This assumes a learning curve where mass-efficiency improves as cumulative sail production increases. The lower bound in this domain similarly assumes $30/kg launch costs. Upper bounds are again 10× the central estimate.

The third domain spans from a South-Pole-$CO_2$-ice-deposit-sized region to "all" of Mars (one hemisphere). For this case, 4 g/m² is still used for spacecraft areal density. But rather than polar orbits, sails are assumed to be located at the Mars-Sun $L_2$ point (displaced 5× closer to the planet due to solar radiation pressure). From this distance, the diameter of the reflected light's spot size is 5× less than the diameter of Mars, meaning no reflected light is lost to spot-size overspill, and the duty cycle of the sail is 100% (compared to ~20% for the South Pole). $10^{16}$ W would need to be intercepted to produce an average of +140 W/m² across the entire antisolar hemisphere of the planet that is visible to a sail at Mars' $L_2$ point. This would require 17 million km² of reflector surface area, or 7×$10^{10}$ kg of mass launched from Earth (assuming 4 g/m²). At $100/kg launch and $100/kg procurement, this corresponds to $14 tn. Should launch costs drop to $30/kg, this becomes $9 tn. In practice this is unaffordable, which motivates sail production at one of Mars' moons (Fig. 21).